\documentclass[twocolumn, pra,showpacs,footinbib,superscriptaddress]{revtex4-1}
\usepackage{graphicx}
\usepackage{color}

\newcommand{\mms}{mm~s$^{-1}$}
\newcommand{\etal}{\emph{et~al.}}

\begin{document}

\title{Superfluid critical velocity of an elongated harmonically trapped Bose-Einstein condensate}
\author{Chao Feng}
\affiliation{School of Mathematics and Physics, The University of Queensland, Brisbane, Queensland 4072, Australia}\author{Matthew J.~Davis}
\email{mdavis@physics.uq.edu.au}
\affiliation{ARC Centre of Excellence in Future Low-Energy Electronics Technologies, School of Mathematics and Physics, The University of Queensland, Brisbane, Queensland 4072, Australia}
\date{\today}

\begin{abstract}
We numerically model experiments on the superfluid critical velocity of an elongated, harmonically trapped  Bose-Einstein condensate as reported by [P.~Engels and C.~Atherton, Phys.\ Rev.\ Lett. \textbf{99},~160405 (2007)].  These experiments swept an obstacle formed by an optical dipole potential through the long axis of the condensate at constant velocity.  Their results  found an increase in the resulting density fluctuations of the condensate above an obstacle velocity of $v\approx 0.3$ \mms, suggestive of a superfluid critical velocity substantially less than the average speed of sound.  However, our analysis shows that the that the experimental observations of Engels and Atherton are in fact consistent with a superfluid critical velocity equal to the local speed of sound.   We construct a model of energy transfer to the system based on the local density approximation to explain the experimental observations, and propose and simulate experiments that sweep potentials through harmonically trapped condensates at a constant fraction of the local speed of sound.  We  find that this leads to a sudden onset of excitations above a critical fraction, in agreement with the Landau criterion for superfluidity. 
\end{abstract}


\maketitle

% --- SECTION 1: INTRODUCTION --- %
\section{Introduction}
Superfluidity is an emergent phenomenon occurring in many distinct physical systems with different underlying microscopic physics~\cite{Schmitt}. One of the defining features of a superfluid is its ability to flow past obstacles below a certain critical velocity $v_c$ without dissipation. Determining what this critical velocity is and how it arises is a key aspect of superfluid research. A phenomenological derivation of the superfluid critical velocity was provided by Landau, based on the dispersion relation of the quasiparticles of the superfluid. By considering how a classical particle can exchange energy and momentum with the superfluid~\cite{Khalatnikov,Annett}, the Landau criterion is
\begin{equation} 
v_c = \mathrm{min}\left[\frac{\epsilon(p)}{p}\right],
\label{eq:landau_criterion}
\end{equation}
where  $\epsilon(p)$ is the energy of a quasiparticle with momentum $p$. 

Perhaps the best known example of a superfluid is helium-4 below the critical temperature of 2.17~K, as first discovered in 1938~\cite{Kapitza, Allen}. The critical velocity in superfluid helium-4 has been measured using a range of methods~\cite{Allum, Castelijns}. When \textit{macroscopic} impurities such as vibrating wires are immersed in superfluid helium, the measured critical velocity for the onset of dissipation is consistently found to be lower than the Landau critical velocity~\cite{Castelijns, McClintock}. However, the situation is different for microscopic obstacles, such as the passage of ions through the superfluid \cite{Allum, Castelijns}. In these experiments the ions experience a drag force above a critical velocity that is in broad agreement with the Landau criterion.

The discrepancy for macroscopic impurities in superfluid helium-4 is not well understood~\cite{McClintock,Barenghi}. Application of the Landau criterion to the excitation spectrum suggests the critical velocity is connected to the formation of a class of excitations called rotons~\cite{Khalatnikov}. However, it is well known that vortices form when superfluid helium-4 is stirred \cite{Hall, Yarmchuk}, and it has been suggested that vortex formation plays a key role in the  determination of the critical velocity \cite{Feynman, McClintock}.

One of the challenges of understanding superfluid helium-4 from a theoretical perspective is that it is a strongly interacting quantum fluid, which poses problems for the development of a microscopic theory. However, the experimental observation of weakly-interacting, dilute gas Bose-Einstein condensates (BECs) in the mid-1990s raised the prospect of predictions for a critical velocity from microscopic theory subsequently being rigorously and quantitatively tested in experiment. The application of Bogoliubov theory~\cite{Pethick} to homogenous dilute gas BECs at zero temperature leads to the well-known Bogoliubov quasiparticle dispersion relation 
\begin{equation}
\epsilon(p) = \left[\frac{p^2}{2m}\left(\frac{p^2}{2m} + 2 g n_0\right)\right]^{1/2},
\end{equation}
where $p$ is the momentum of the quasiparticle, $m$ is the particle mass, $g=4\pi \hbar^2 a /m$ is the strength of particle interactions characterised by the $s$-wave scattering length $a$, and $n_0$ is the condensate density. Applying the Landau criterion to  BECs with repulsive interactions $a>0$ suggests that at zero temperature they are superfluids with a critical velocity equal to the speed of sound $c_s = (2 g n_0/m)^{1/2}$.  

However, a direct comparison between the Bogoliubov prediction for $v_c$ and experimental results is complicated by the fact that until recently the majority of quantum gas experiments are performed with the atoms confined by some form of harmonic trap.  This leads to BECs with inhomogeneous density profiles at equilibrium, and as the local speed of sound depends on density,  the critical velocity should be a function of position.

Despite this complication, there have been several experiments that have probed the critical velocity of trapped BECs~\cite{Raman, Onofrio, Engels, Neely, Ramanathan,Desbuquois,Moritz}. Typically in these experiments, the optical dipole potential of a focussed laser beam forms the impurity which is moved within the BEC. The first experiment was by Raman \emph{et al.}~\cite{Raman}, followed closely by Onofrio \emph{et al.}~\cite{Onofrio} in which a tightly focussed potential was sinusoidally moved within a prolate (cigar-shaped) BEC.  The heating of the condensate was measured as a function of the frequency of oscillation, and the results showed an apparent threshold frequency for the onset of heating corresponding to a speed of the obstacle significantly lower than the speed of sound. A limitation of this experimental procedure is that the oscillation of the potential causes the impurity to backtrack through its own wake where the density has been disturbed. 

A later experiment by Engels and Atherton in 2007~\cite{Engels}, the focus of this paper, overcame the wake problem by sweeping the potential formed by a laser sheet through a prolate BEC in a single pass. When the speed of the laser exceeded an apparent threshold, dark density bands were observed to form  in its wake. Again, the threshold velocity above which excitations were observed was lower than estimates for the critical velocity based on the average speed of sound. 

In 2010 an experiment by Neely~\emph{et al.}~\cite{Neely} stirred an oblate (pancake-shaped) BEC using a focussed laser beam.  The potential began at rest within the BEC, and was then moved at constant velocity through the centre, before the intensity was ramped to zero as it approached the edge of the system. It was found that when the laser moved above a threshold speed vortex dipoles formed in its wake.  Again, the threshold for vortex dipole formation was found to be lower than the speed of sound at the centre of the BEC.

In 2011 Ramanathan~\emph{et al.} performed experiments with a toroidal (ring-shaped) BEC that initially had a single unit of circulation~\cite{Ramanathan}. They inserted a stationary barrier and measured the survival probability of the phase winding for different barrier heights.  They observed a critical barrier height above which there was a significant probability of current decay. By computing the local speed of sound $c_l$ and flow velocity $v$ from the column density, they also found the ratio $v/c_l$ to be significantly less than one.

An experiment by Desbuquois~\emph{et al.} in 2012~\cite{Desbuquois} stirred a disk-shaped 2D Bose gas in its superfluid phase. The stirring was performed in a circular trajectory about the centre of the gas so that the density along the stirrer's path was approximately constant. The resulting change in temperature was measured as a function of the stirring speed, and indicated a threshold velocity for heating that was smaller than the local speed of sound along the stirring path~\cite{Desbuquois}.

Weimer~\emph{et al.} have performed similar experiments to Desbuquois~\emph{et al.}~\cite{Desbuquois} in a superfluid Fermi gas to measure the critical velocity for the BEC-BCS crossover~\cite{Moritz}.  They used an attractive stirring obstacle moving in a circular path in an oblate harmonic trap for a fixed time, and looked for heating as a function of the velocity of the obstacle.  A sudden increase in the heating rate provided evidence for a critical velocity, which was found to be consistently lower than the measured speed of sound.  However, in the BEC regime they modelled their experiments using classical field simulations, and found good agreement with the observations after accounting for the effects of finite temperature and the circular stirring path through inhomogeneous density, which had individual contributions to the reduction in $v_c$ of about 15\%.

Theoretical insight into the superfluid critical velocity has been found from analytical and numerical studies of the Gross-Pitaevskii equation, a mean-field equation of motion for the condensate order parameter. In two and three dimensions, numerical studies have shown that the motion of localised potentials through BECs leads to the production of vortices and sound waves above a threshold velocity~\cite{Jackson_Vortex, Jackson_Simulate_Raman, Frisch, Winiecki_Vortex}.  In one-dimensional homogeneous systems,  the motion of a localised potential through a BEC results in the formation of grey solitons above a threshold velocity~\cite{Hakim, Astrakharchik}.  Analytical studies in such systems \cite{Watanabe, Leboeuf, Hakim, Pavloff} have shown that a moving impurity of vanishing size and depth (i.e.\ a microscopic impurity) creates excitations at precisely the speed predicted by the Landau criterion Eq.~(\ref{eq:landau_criterion}). However, an impurity with depth $V_0$ that is a moderate fraction of the system chemical potential $\mu$ lowers the critical velocity below that predicted by the Landau criterion, as a result of increasing the local fluid flow and decreasing the local density \cite{Hakim, Pavloff, Leboeuf, Watanabe}. Despite a significant amount of work in this area, there is yet to be a satisfactory reconciliation between theoretical predictions of the critical velocity arising from investigations of the homogeneous GPE with a repulsive obstacle and observed values from experiments. This enduring discrepancy is a key motivation for this paper.  

In specific cases, the critical velocity can be estimated by assuming an explicit form for the condensate wave function after it has been stirred~\cite{Neely, Ramanathan, Moulder}. This allows the calculation of the associated energy and momentum change, which can be used to estimate the critical velocity using the Landau criterion~Eq.~(\ref{eq:landau_criterion}). This \emph{a posteriori} approach was used to estimate the critical velocity in Ref.~\cite{Neely}, where a vortex pair forms when the impurity exceeds the critical velocity~\cite{Crescimanno}. 

Elsewhere, numerical simulations have been performed with the aim of understanding experimental results.  Findings from Piazza~\emph{et al.} indicate a more sophisticated understanding of the local density and flow conditions is necessary~\cite{Piazza} to accurately model the experiment of Ramanathan~\etal~\cite{Ramanathan}, while Mathey \emph{et al.}~find that finite temperature thermal fluctuations could lead to excitations below the mean-field critical velocity~\cite{Mathey} for the conditions reported in the same experiment. While these works provided useful insights, neither provided conclusive reconciliation between theoretical predictions of the critical velocity and experimental observations.
 
There remains enduring interest in further understanding the theory-experiment discrepancy. In this paper we model the experiment of Engels and Atherton~\cite{Engels} using dynamical three-dimensional simulations of the zero temperature Gross-Pitaevskii equation (GPE) with no free parameters, and use these to interpret the experimental observations. We will show such experiments are not conducive to measuring a single critical velocity and in fact, reflect a spatially dependent critical velocity as the obstacle traverses a BEC with an inhomogeneous density profile.  We then focus on understanding how energy is transferred to the BEC by the creation of solitonic excitations above the critical velocity. This provides a cumulative measure of how a spatially dependent critical velocity manifests in such systems. We  then develop a simple one-dimensional model that broadly reproduces the 3D GPE results. We demonstrate how the Landau criterion as applied to determine the critical velocity in homogeneous systems can be used to quantitatively predict the observed energy transfer. Finally, we describe a scheme to measure the effective homogeneous superfluid critical velocity in a prolate harmonically trapped BECs, and find that the same procedure is also effective for harmonically trapped BECs in oblate traps. 

% --- SECTION 2: MODEL --- %
\section{Model}

% --- SUB-SECTION: Summary of Engels and Atherton experiment --- %
\subsection{Summary of Engels and Atherton experiment}
The simulations presented in this paper are directly related to the experiments performed by Engels and Atherton~\cite{Engels}.  For the convenience of the reader we provide a summary of the experimental procedure and results.  Engels and Atherton worked with a elongated harmonically trapped BEC of $^{87}$Rb atoms in the $F=1, m_F = -1$ hyperfine state. The trap had a prolate geometry with frequencies $(\omega_x,\omega_r)/2\pi  =  (7.1,203)$ Hz for the axial and radial directions respectively.  The experiments began with $N=4.5\times10^5$ atoms  and no discernible thermal cloud, resulting in a chemical potential of $\mu = 277 \hbar \omega_x$, a peak density of $n = 2.5 \times 10^{14}$~atoms~cm$^{-3}$, and radial and axial Thomas-Fermi lengths of 6.6 and 195 $\mu$m respectively.

The BEC was probed by sweeping a Gaussian optical dipole potential modelled as $U(x,t)$ (the ``obstacle'') at constant speed $v$ through the BEC
\begin{equation}
U(x,t)=U_{0}\exp\left[-2(x + vt - x_0)^{2}\sigma^{2}\right],
\label{eq:gaussian_obstacle}
\end{equation}
where $x_0$ is the position of the obstacle at $t=0$, $\sigma$ is the characteristic width, and $U_0$ is the height.  For all experiments the obstacle width was $\sigma = 7.6$ $\mu$m~\cite{Engels}.  For comparison, the healing length of the BEC at the centre of the trap is $\xi = (\hbar^2/2mng)^{1/2} = 0.17$ $\mu$m $ \ll \sigma.$ The obstacle is assumed to be homogeneous in the other dimensions ($y, z$) where the width of the obstacle is greater than the condensate span. The obstacle began outside the BEC, and stopped at a final position approximately three-quarters through it.  The atoms were immediately transferred from a trapped state to an anti-trapped state to rapidly expand the condensate in 3~ms of time of flight before absorption imaging.

Several sets of experiments were performed with both repulsive and attractive obstacles.  In this paper we focus on the results reported with a repulsive obstacle of a height of $U_0= 0.24 \mu$.  The range of obstacle speeds considered was from $v=0.25$~\mms\ to $v=3.3$~\mms, compared with the speed of sound at the centre of the BEC of $c\approx 3.0$~\mms, or $2.1$~\mms\  if averaged across the radial dimension~\cite{Engels}. 

% -- FIGURE 1 -- %
\begin{figure}
\includegraphics[width=\columnwidth]{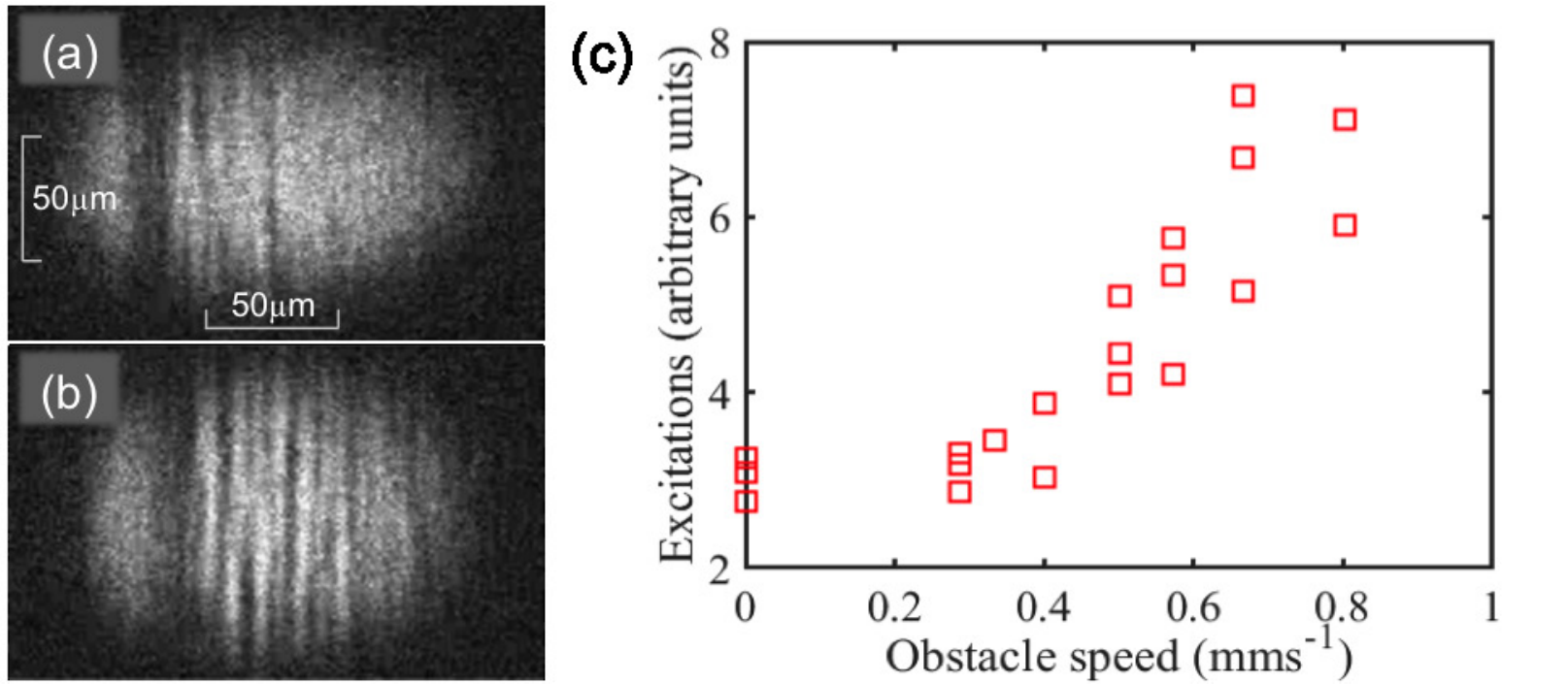}
\caption{Reproduction of results from experiment in Ref.~\cite{Engels} with scalebar added. (a) and (b) show condensate images after the passage of the obstacle at speeds of 0.5 and 0.7~\mms\ respectively.  (c) is a reproduction of the results of the excitation measure. Reprinted Figure 1 (d) and (f) with permission from Ref.~\cite{Engels}. Copyright 2019 by the American Physical Society.\label{fig:typical_exp}}
\end{figure}
% -- FIGURE 1 -- %

The experimental procedure described above often resulted in a number of stripe-like excitations in the density of the BEC. A typical result is reproduced in Fig.~\ref{fig:typical_exp}(a,b). The amount of excitation depended on the speed of the obstacle, and was quantified by calculating the root-mean-square (RMS) of the difference between a Fourier-smoothed and the original absorption image of the central region of the condensate at the end of the experimental run~\footnote{The experimental sampling region was 35 $\mu$m in the $x$-direction by 70 $\mu$m in the $z$-direction after the anti-trapping expansion.}.  However, this analysis was not performed for obstacle speeds higher than $v = 0.8$~\mms\ as ``the broad low-density region that develops in the wake of the barrier affects our measure''~\cite{Engels}.

The results of this analysis are compared to the obstacle speed, and are reproduced in Fig.~\ref{fig:typical_exp}(c). The data indicates that there is little excitation of the condensate for obstacle speeds below $v \sim 0.3$~\mms, and this is identified as the critical velocity for excitations in this experiment~\cite{Engels}. 

%  --- SUB-SECTION: Model --- %
\subsection{Model}
To analyse these experimental results we perform simulations of the time-dependent Gross-Pitaevskii equation for the trapped BEC~\cite{Gross, Pitaevskii} using the XMDS numerical package \cite{Dennis}.  The GPE has proved remarkably successful in quantitatively modelling a variety of experiments on BECs, and has been used to model a number of scenarios probing the superfluid nature of the system, for example Refs.~\cite{Hakim,Astrakharchik, Jackson_Vortex,Jackson_Simulate_Raman, Frisch, Carretero-Gonzalez, Winiecki,Radouani, Piazza, Jackson_Simulate_Raman}. 

The harmonic trap in which the experiments were performed possess cylindrical symmetry, and this can be utilised to reduce the number of dimensions in the model. We define the dimensionless space and time coordinates as $x'=x/\sqrt{\hbar/m\omega_x}$, and  $t'=\omega_x t$.  Dropping dashes for convenience, the resulting Gross-Pitaevskii equation in cylindrical coordinates is 
\begin{eqnarray}
i \frac{\partial}{\partial t}\psi(r,x)  &=& \left[ -\frac{1}{2}\left(\frac{\partial^{2}}{\partial r^{2}}+\frac{1}{r}\frac{\partial}{\partial r}+\frac{\partial^{2}}{\partial x^{2}}\right) +U(x,t) \right. \nonumber\\
&  + &\frac{1}{2} \left(r^{2}+\gamma_{x}^{2}x^{2}\right) \left.\vphantom{\frac{1}{1}} + g|\psi(r,x)|^{2}\right]\psi(r,x),
\label{eq:GP}
\end{eqnarray}
where $1/\gamma_{x}=\omega_{r}/\omega_{x}\approx29$ is the aspect ratio of the harmonic trap, and the dimensionless interaction constant is $g=4\pi a/\sqrt{\hbar/m\omega_x}$, where $a$ is the s-wave scattering length for $^{87}$Rb. The obstacle potential $U(x,t)$ is the dimensionless form of Eq.~(\ref{eq:gaussian_obstacle}).  The speed of the obstacle  $v$ is constant for each simulation.

We note a limitation of the cylindrical GPE is that it formally preserves rotational invariance. While this is consistent with the assumption that the obstacle is effectively 1D, it may exclude any potential symmetry breaking excitation or decay mechanism that may be available in the experimental situation, for example, arising from quantum or thermal fluctuations. We will note in the paper where such effects may be relevant, but a detailed investigation is outside the scope of this work.

% --- SECTION 3: RESULTS OF GP SIMULATIONS --- %
\section{Results of Gross-Pitaevskii simulations}

% --- SUB-SECTION: Comparison of column density images --- %
\subsection{Comparison of column density images}
\begin{figure}
\includegraphics[width=\columnwidth]{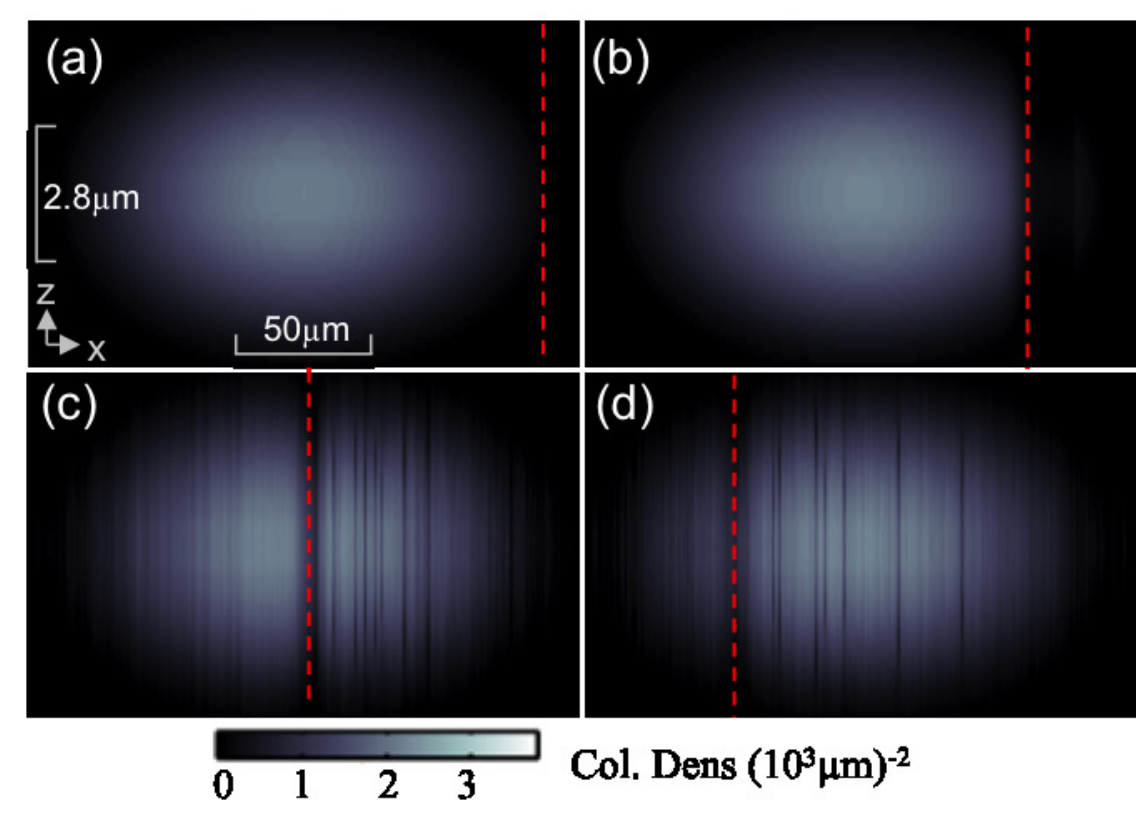}
\caption{Example condensate column densities from a typical 3D GPE simulation as described in the text with obstacle speed $v=0.5$~\mms. Note the different length scales in the $z$ and $x$ dimensions. (a) Initial state before the obstacle has entered the BEC. (b--d) BEC column density at early ($t=116$~ms), middle ($t=233$~ms) and final ($t=329$~ms) times respectively. The vertical red dashed line in each frame indicates the position of the centre of the obstacle.\label{fig:typical_simulation}}
\end{figure}

We have performed full simulations of the experiment for a range of obstacle velocities from $v=0.06$~\mms\ to $v=6.0$~\mms.  Several examples of \emph{in situ} column densities of the condensate for a typical simulation with  an obstacle speed of $v=0.5$~\mms\ are shown in Fig.~\ref{fig:typical_simulation}.

% -- FIGURE 2 -- %
% -- FIGURE 2 -- %

The obstacle enters the BEC from the right hand side, displacing the condensate to create a dark density depression. By the time the obstacle is in the centre of the system [Fig.~\ref{fig:typical_simulation}(c)] it has generated several excitations in its wake, seen as thin, dark bands.  Examining the phase of the BEC wave function (not shown) shows that these excitations are a mixture of solitary waves, ring vortices and dispersive waves. Solitary waves are identified by the presence of $\pi$ phase steps across the axial dimension, and are  distinguished from dispersive waves by their extended survival time when the simulations are run significantly beyond the experimental timeframe.  In Fig.~\ref{fig:typical_simulation}(d) the obstacle has reached its final position, approximately three-quarters of the way through the BEC. 

A region of increased density is also observed to form to the front of the obstacle which remains only when the obstacle is moving.  This feature appears to be hydrodynamic in nature and has been previously observed in theoretical studies of 1D GPEs~\cite{Leszczyszyn,Pavloff,Hakim}. A small amount of dispersive shock waves are also generated in front of the obstacle when it first enters the condensate.  Both of these features are evident to the left of the obstacle in Fig.~\ref{fig:typical_simulation}(c).

\begin{figure}
\includegraphics[width=0.95\columnwidth]{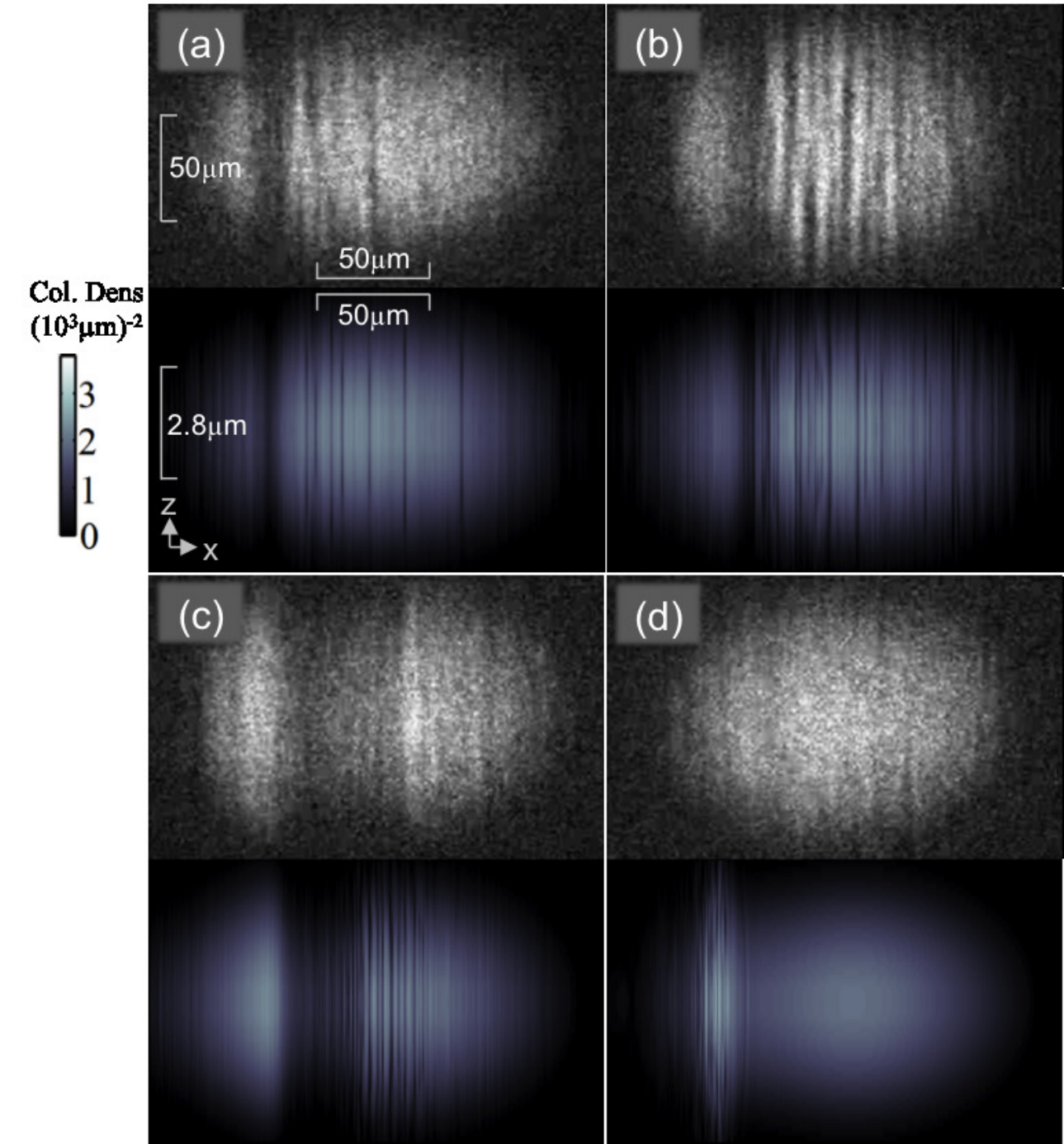}
\caption{Comparison of condensate column densities following anti-trapping time-of-flight expansion from the experiment of Engels and Atherton~\cite{Engels} (top) with the \emph{in situ} column densities from the GPE simulations (bottom) for a range of obstacle velocities $v$, with scalebars added to the original image. (a) $v = 0.5$~\mms.  (b) $v = 0.7$~\mms. (c) $v = 1.3$~\mms. (d) $v = 3.5$~\mms. Reprinted Figure 1 (d), (f), (i), (k) with permission from Ref.~\cite{Engels}. Copyright 2019 by the American Physical Society.
\label{fig:exp_absorption}}
\end{figure}

Figure~\ref{fig:exp_absorption} presents examples of \emph{in situ} column densities at the time when the obstacle has reached its final position for a range of obstacle speeds from $v=0.5$~\mms\ to $v=3.5$~\mms. These are compared to the corresponding experimental absorption images taken following rapid expansion in a repulsive harmonic potential.  As the expansion is not simulated, these images are not directly comparable. However, we note the broad agreement between the simulation images and the experimental data, and that density variations in the experimental images occur  mostly along the $x$-dimension. These are supportive of our assumption that the main effect of the anti-trapping is to expand the condensate in the radial dimension.

Both the experimental and simulation data show the generation of more excitations in the wake of the obstacle for faster obstacle speeds [Fig.~\ref{fig:exp_absorption}(a,b)], and the formation of a wide band of low density at particular obstacle speeds [Fig.~\ref{fig:exp_absorption}(c)].  Examination of the GPE wave function in this case reveals that this low density band is actually a cluster of tightly packed dark solitary waves and vortex rings which would not be resolvable in the experimental imaging.  We will address the cause of the clustering effect in a later section. The simulation dynamics show that the vortex rings arise due to the decay of solitary waves generated in the wake of the obstacle. We note the possibility there may be symmetry-breaking decay processes available in the experiment that the cylindrical GPE would not capture due to its rotational invariance.

Figure~\ref{fig:exp_absorption}(d) suggests a suppression of excitations in the wake of the obstacle at higher obstacle speeds. This feature of suppressed excitations when condensates are stirred with fast obstacles has previously been discussed in theoretical work in 1D~\cite{Radouani, Pavloff, Leboeuf}. In our simulations, we find that the speed of solitary waves is related to the speed of the obstacle that created it. At lower obstacle velocities the solitary wave it creates is very dark --- having near zero density and little or no initial velocity of its own. As the obstacle speed increases, the resultant solitary wave is less dark --- its speed increases while its amplitude decreases. This  observation is consistent with the established behaviour of solitons in 1D~\cite{Tsuzuki}. As the obstacle speed increases further, it creates progressively less dark solitary waves until it ceases to produce solitary waves at all.

Notably, the increased density region in front of a moving obstacle continues to appear at higher velocities. This is clearly visible in the \emph{in situ} simulation image in Fig.~\ref{fig:exp_absorption}(d) (bottom). The feature is not visible in the corresponding experimental image Fig.~\ref{fig:exp_absorption}(d) (top). This is likely because the feature, being hydrodynamic in nature, disperses rapidly once the obstacle stops moving, as in the case for the experimental image which was taken after 3~ms of expansion.

% -- FIGURE 3 -- %
% -- FIGURE 3 -- %

% --- SUB-SECTION: Comparison with Experiment --- %
\subsection{Comparison with excitation measurements}
\begin{figure}
\includegraphics[width=0.82\columnwidth]{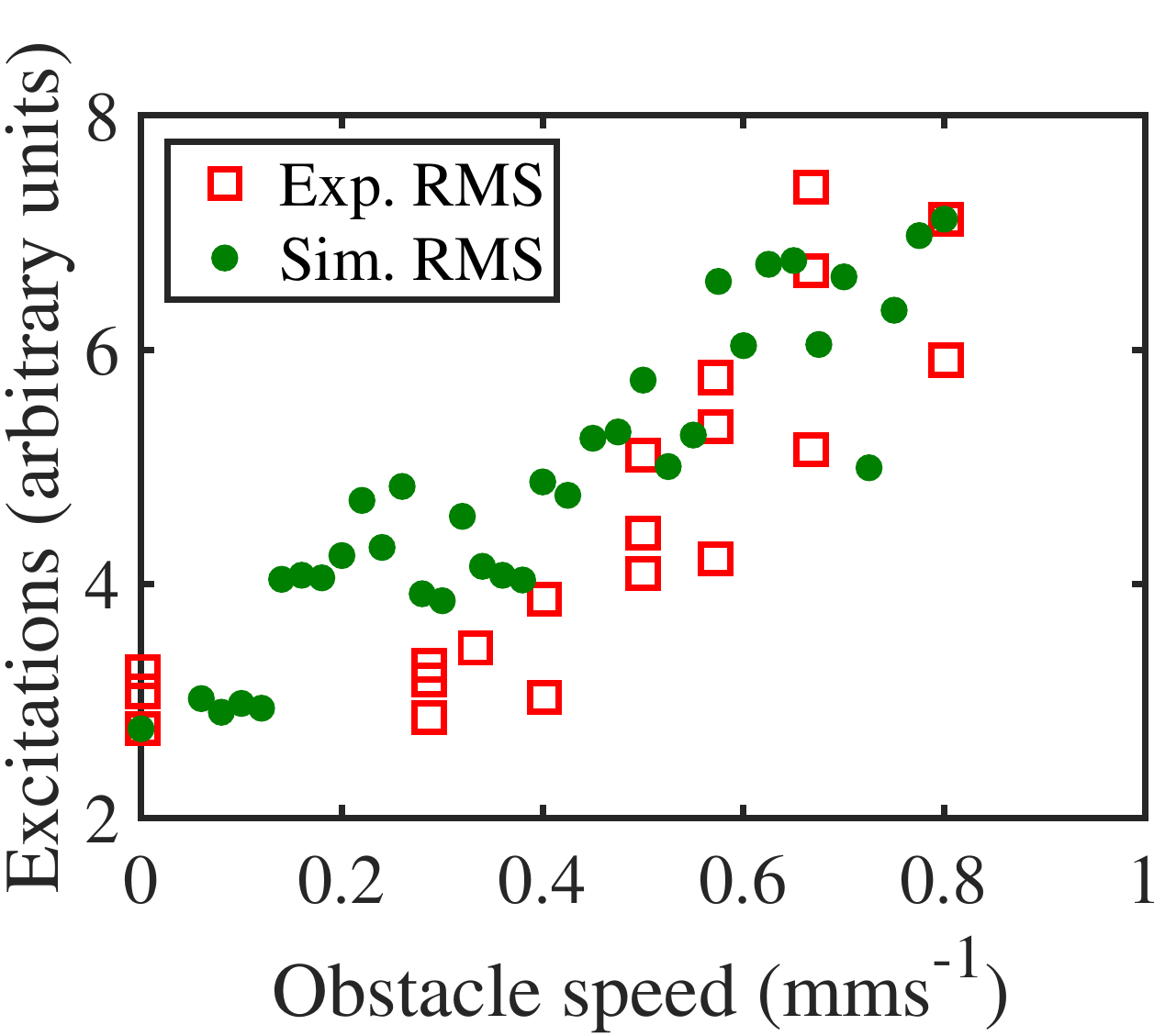}
\caption{Comparison of the experimental RMS measure of condensate excitations  (green dots)~\cite{Engels}  with the same quantity reconstructed from the 3D GPE simulations (red squares) as a function of obstacle velocity, as  described in the text.  The simulation results  were scaled on the vertical axis to match the experimental data at $v= 0$~\mms\ and $v= 0.8$~\mms.
\label{fig:exp_RMS}}
\end{figure}

To provide an indicative comparison with the experimental measure of excitations, we calculate the sum of the root mean square deviation between the column densities of the unperturbed ground state and perturbed system following passage of the obstacle~\footnote{We did not attempt to match the Fourier analysis method of the experiment to obtain the ground state density profile because (i) it would introduce further arbitrary scaling and conversion parameters and (ii) in our simulations, we have access to the unperturbed ground state density profile}. As in the experiment, only the central region is considered~\footnote{The experiment stated a central window of 35~$\mu$m in the $x$-direction and 70~$\mu$m in the $z$-direction based on images obtained after expansion. To estimate the effects of the expansion, we used the experimental images to calculate the ratio of the window size and the span of the condensate in the $z$-direction, and applied the same ratio to our simulated images, giving around 0.5~$\mu$m in the $z$-direction.}.  The experimental results are reproduced in~Fig. \ref{fig:exp_RMS} as red squares, and the simulation results are plotted as green dots. The experimental results were reported in arbitrary units, and so we have scaled and centred our simulated RMS data along the vertical axis to match the range of experimentally reported values ($v=0$~\mms\ to $v=0.8$~\mms).   We therefore caution that it does not represent a direct quantitative comparison, and is only indicative of the trend.  Despite this, the comparison shows broadly similar features --- both sets of data are suggestive of a transition from little or no excitations to many excitations above a certain obstacle speed. However for the simulations this is closer to $v \approx 0.15$~\mms\ as compared to $v \approx 0.3$~\mms\ for the experiment.  In comparison, the local speed of sound determined from the density at the centre of the trapped BEC is approximately an order of magnitude larger, as in the experiment~\cite{Engels}.

However, close inspection of the GPE simulations makes it clear that this simple interpretation of the data is not correct.  Towards the boundary of the system the density of the condensate approaches zero. This means that in the local density approximation (LDA) the local speed of sound (and hence critical velocity) also tends to zero.  A moving obstacle can be expected to generate excitations in these regions at almost arbitrarily low speeds~\footnote{While the critical velocity is expected to vanish, finite size effects will impose a minimum threshold unrelated to Landau's critical velocity and is not consequential to our present discussion.}. Indeed, even simulations at the lowest simulated obstacle speed of $v = 0.06$~\mms\ show the formation of solitary waves and dispersive waves at the edge of the system. When a low speed obstacle moves further into the condensate, away from the low density regions, it no longer exceeds the local critical velocity and correspondingly, ceases to generate excitations. In general, higher obstacle speeds are able to create excitations over a greater spatial region, thus, generating more excitations during their passage.

An important observation is that solitary waves created in the low density outer regions of the condensate move to be in the centre of the condensate at the time of measurement.  A solitary wave created in isolation would oscillate freely in the trapped condensate, much like a particle in a well (see for example, Ref.~\cite{Frantzeskakis}). However, in this case, the motion of the solitary wave is impeded by the moving obstacle, with solitary waves clustering in its wake until it passes the condensate centre. This transfers most of the density depressions to the central region of the condensate, which can then contribute to the experimental RMS measure of excitations for obstacle speeds that do not actually create solitary waves in the central region. After the obstacle has passed the centre, the density depressions continue to move in the condensate and are largely responsible for the noise in the simulated RMS data --- the RMS measure is noticeably affected by a single solitary wave being inside or outside the measurement window.

Whether by good planning or good fortune,  our simulations indicate that the point at which the images are taken, corresponding to when the obstacle has travelled approximately three-quarters of the way through the system, occurs at a time when a large fraction of the excitations formed in the wake of the obstacle are clustered in the central sampling region of the condensate.  The RMS density measurement performed in the experiment is therefore likely to be a reasonable indication of total excitations to the condensate at low velocities. If it had been applied at higher velocities, where the formation of solitary waves are more copious and occur throughout the condensate, it would have been a less robust measure.

% -- FIGURE 4 -- %
% -- FIGURE 4 -- %

\subsection{Energy transfer to condensate}
\begin{figure}
\includegraphics[width=0.82\columnwidth]{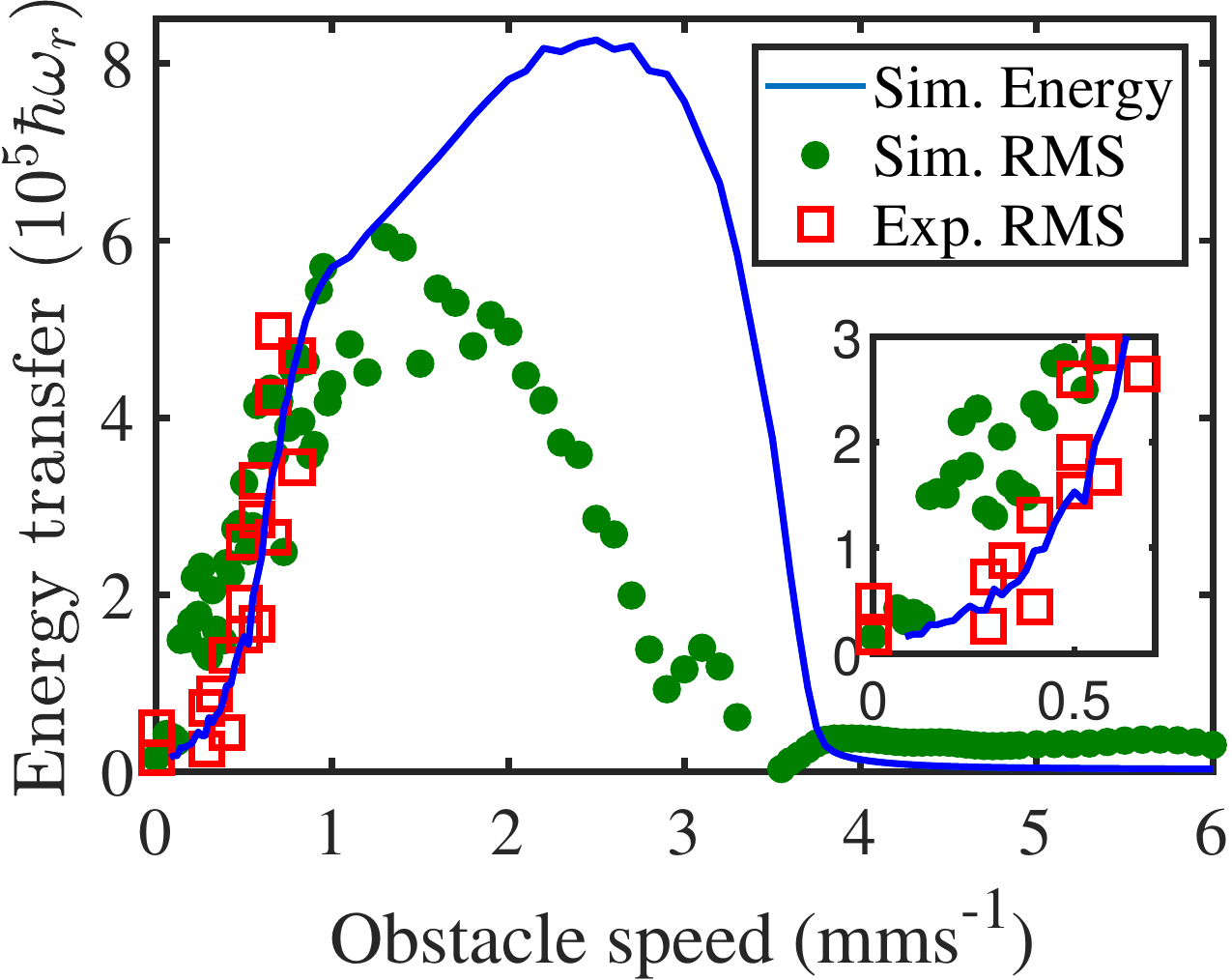}
\caption{Energy transferred to the condensate in 3D GPE simulations  as a function of obstacle speed (blue line), compared with the RMS measures of excitation (red squares: experimental data, green dots: GPE simulations).  The RMS measure (in arbitrary units) has been scaled to best fit the energy transfer curve over the range $v=0$~\mms\ to $v=0.8$~\mms. The inset is a zoomed-in view for $v = 0$~\mms\ to $v=0.7$~\mms.  We note that the energy of the condensate increased for all obstacle velocities. \label{fig:exp_energy}}
\end{figure}

From the simulations the increase in energy of the condensate can be calculated directly. The total energy of the BEC at time $t$ is
\begin{eqnarray}
E(t)&=&\int d\mathbf{r} \,\psi^{*}(\mathbf{r},t)\left[-\frac{1}{2}\nabla^{2}
 + \frac{1}{2}(r^{2}+\gamma_{x}^{2}x^{2})
 \right. \nonumber\\
& & \left. + U(x,t) + \frac{g}{2}|\psi(\mathbf{r},t)|^{2}\right]\psi(\mathbf{r},t).
\label{eq:Energy_integral}
\end{eqnarray}
We calculate the energy transferred to the condensate as
\begin{equation}
\Delta E = E(t_{f}) - E(t_{0}).
\label{eq:energy_transfer}
\end{equation}
as a function of obstacle velocity $v$, and this is plotted as the solid blue line in Fig.~\ref{fig:exp_energy}.  It is compared to both the experimental (red squares) and simulated (green dots) RMS excitation measure, whose arbitrary units have been scaled and centred along the vertical axis (as before) to give the best agreement with this quantitative measure. The results confirm our earlier qualitative observations, that for the range of experimental data presented, the RMS excitation measure appears to give a reasonably good indication of the relative magnitude of energy transfer to the BEC at different velocities. However, for larger obstacle velocities, the simulated RMS measure indicates a faster fall off in energy transfer compared to the direct energy calculation --- faster than the expected fall off from the high velocity suppression effect discussed earlier.  The explanation for this is as follows: for larger obstacle velocities solitary waves are created with faster speeds, smaller amplitudes and broader widths~\cite{Tsuzuki}. These `smeared out' solitary waves lead to a smaller RMS measure.
Furthermore, a greater number of solitary waves are created such that they do not all accumulate in the central observation window, and are thus not accounted for.

The energy transfer curve in Fig.~\ref{fig:exp_energy} illustrates several features of this experiment.  Firstly, and most importantly, it is clear there is no sharp transition in this quantity indicating a single critical velocity for superfluidity.  Secondly, the amount of energy transferred to the BEC initially increases for increasing obstacle speeds, as might be expected as faster obstacles exceed the local critical velocity in greater portions of the condensate. Thirdly, at larger obstacle speeds (beyond the peak local critical velocity of the system) it is demonstrated that excitations are suppressed. Similar behaviour has previously been observed in theoretical investigations of the 1D homogeneous GPE~\cite{Pavloff, Leboeuf, Hakim, Radouani}. We note for completeness that the experiment also considered the case of an attractive obstacle potential, for which RMS data was collected for a broader range of velocities. The experiment observed a similar suppression of excitations at higher velocities. Further exploration of the attractive potential is outside the scope of this work.

% -- FIGURE 5 -- %
% -- FIGURE 5 -- %

% --- SECTION 4: ENERGY TRANSFER --- %
\section{Model of energy transfer for trapped condensates}
The GPE simulations  demonstrate that the obstacle velocity below which excitations were observed to be suppressed in the experiment of Engels and Atherton is not simply related to the Landau criterion.  Instead, excitations are created for most obstacle velocities, and larger obstacle speeds result in greater energy transfer to the condensate up to about $v \approx 2.5$~\mms, before the energy transfer decreases close to zero for obstacle speeds greater than $v \approx 4.0$~\mms. 

A qualitative observation from the GPE simulations is that for a given  speed, the obstacle initially generates excitations in the low density outer region of the condensate, before it moves without drag through the high density inner region. Conversely, at sufficiently high obstacle speeds, excitations are first suppressed in the low density outer regions, while excitations occur in the higher density central regions. The spatial boundary between these two types of response occurs at higher densities for faster obstacle speeds. This is suggestive of a critical velocity that depends sensitively on the density along the path of the obstacle.  

The simulations suggest the experiment was mostly successful in limiting dynamics to 1D. Due to the large aspect ratio of the harmonic trap and the fact that the obstacle varies only in the axial dimension, most of the excitations are created through axial dynamics, and their spatial variation is only in the axial dimension. The only exception are vortex rings, but these are produced from the decay of solitons rather than as a direct product of perturbation by the obstacle. This is in contrast to other experiments on frictionless flow in BECs which have worked with tightly focussed laser beams resulting in two-dimensional Gaussian obstacles~\cite{Raman,Onofrio,Neely}.

These observations have led us to construct an effective one-dimensional model of the experiment, and make use of the local density approximation in order to utilise known results for moving impurities in the 1D homogeneous GPE~\cite{Hakim, Pavloff, Watanabe}. As the obstacle in the experimental system experiences a variety of critical velocities rather than a single threshold, we consider the energy transfer as a cumulative measure of the critical velocities in the path of the obstacle.  We do this by first identifying regions of superfluidity for a given obstacle speed, drawing on previous work in 1D homogeneous systems. We introduce a simple model to describe the energy transfer at a given density and integrate over the condensate density profile to predict the total energy transfer.  The model is described in detail below, and allows us to connect the GPE results to the Landau criterion for the critical velocity.
 
% --- SUB-SECTION: 1D Representations --- %
\subsection{One-dimensional approximation}
The most straightforward dimensional reduction of the GPE to 1D assumes a product form for the wave function $\psi(r,x) = \psi(x) \phi(r)$, with
\begin{equation}
\phi(r) \propto \exp\left(-\frac{r^2}{2\sigma_r}\right),
\label{eq:gaussian_profile}
\end{equation}
where $\sigma_r$ is a static parameter that characterises the width of the radial wave function. Integrating out the radial dimension results in the reduced GPE (RGPE)
% EQUATION
\begin{eqnarray}
i\hbar\frac{\partial}{\partial t}\psi(x,t) &=& \left( -\frac{1}{2}\frac{\partial^{2}}{\partial x^{2}} + \frac{1}{2}\omega_{x}^{2}x^{2}+U(x,t) \right.\nonumber\\
& & \left.\vphantom{\frac{1}{1}} + g_{\rm 1D}|\psi(x,t)|^{2}\right)\psi(x,t),
\label{eq:RGP}
\end{eqnarray}
% EQUATION
where $U(x,t)$ has the same form as in Eq.~(\ref{eq:gaussian_obstacle}).  In order to represent the full 3D system, one could choose the chemical potential to give the same Thomas-Fermi length for the axial direction~\cite{chiofalo}, which results in an effective 1D interaction parameter
\begin{equation}
g_{1D} = \frac{2(2\mu)^{3/2}}{3\omega_x/\omega_0 N}.
\end{equation}

\begin{figure}
\includegraphics[width=0.82\columnwidth]{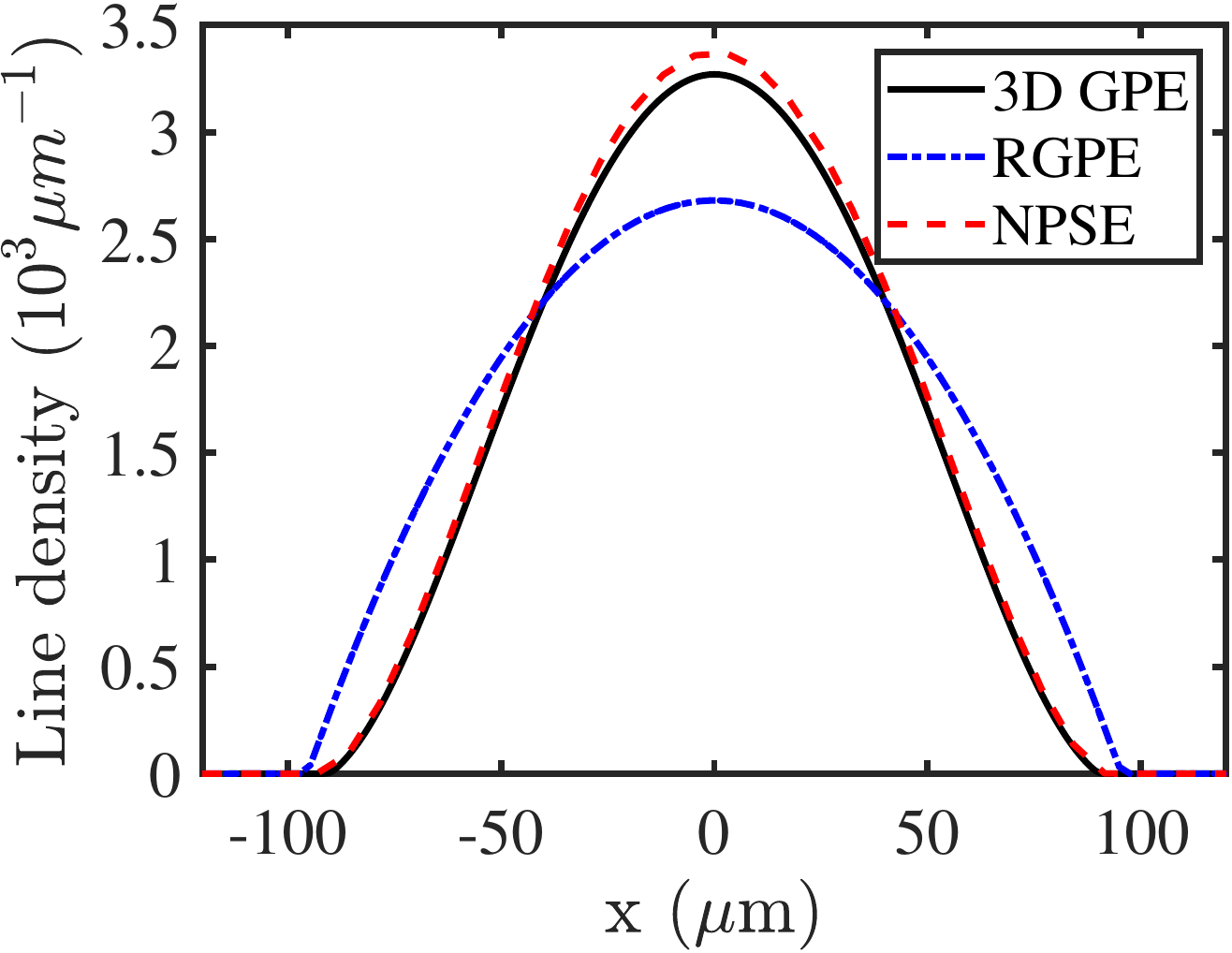}
\caption{Comparison of the 1D line density of the ground state solution of the 3D GPE (black solid line) with two 1D approximations --- the RGPE (blue dot-dashed line) and the NPSE (red dashed line) for the experimental parameters of Ref.~\cite{Engels}.
\label{fig:1d_groundstates}}
\end{figure}

However, it has previously been shown that the ground state density profile obtained using this reduction does not  agree well with the integrated ground state profile of the full GPE when interactions broaden the density profile in the radial dimension~\cite{Salasnich}.  An improved approximation to the full 3D GPE for this situation can be obtained by allowing  the width of the assumed Gaussian profile in the radial dimension to vary along the axial dimension and in time $\sigma_x = \sigma_r(x,t)$. A 1D equation of motion for the wave function in the remaining dimension $\psi(x, t)$ can be found through integration of the associated action integral (for details, see \cite{Salasnich_cigar, Salasnich}), leading to the so-called non-polynomial Schr\"{o}dinger equation (NPSE)
% EQUATION
\begin{eqnarray}
&i&\frac{\partial }{\partial t} \psi(x,t) =\left(-\frac{1}{2}\frac{\partial^{2}}{\partial x^{2}}+\frac{1}{2} x^{2}+U(x,t)\right.\nonumber\\
& & \left. + \frac{1}{2}\sqrt{1+\alpha|\psi(x,t)|^{2}} + \frac{1/2 + \alpha|\psi(x,t)|^{2}}{\sqrt{1+\alpha|\psi(x,t)|^{2}}}\right)\psi(x,t).\nonumber\\
\label{eq:NPSE}
\end{eqnarray}
% EQUATION
where $\alpha=2a_s/x_0$ and  $U(x,t)$ is as in Eq.~(\ref{eq:gaussian_obstacle}). This 1D reduction provides a much closer representation of the integrated density of the 3D GPE ground state. This is demonstrated in Fig.~\ref{fig:1d_groundstates} where we compare ground state line density obtained from the RGPE and NPSE against the integrated line density obtained from the 3D GPE. Note that the NPSE and RGPE do not only differ in the resulting ground state density, but the NPSE can reflect additonal dynamical properties not accessible with the RGPE.

% -- FIGURE 6 -- %
% -- FIGURE 6 -- %

\begin{figure}
\includegraphics[width=0.82\columnwidth]{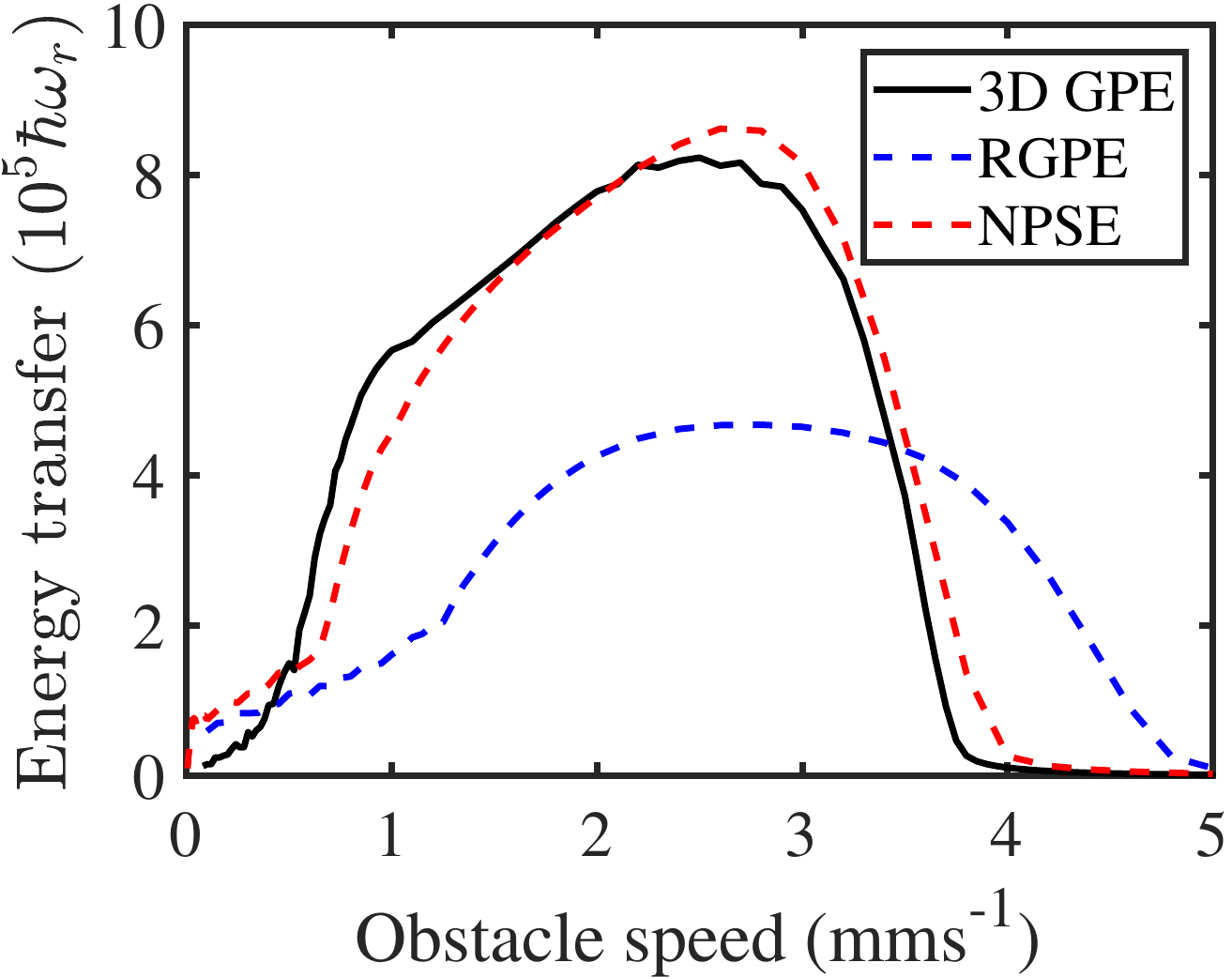}
\caption{Energy transfer (Eq.~\ref{eq:energy_transfer}) as a function of obstacle speed for simulations of the 3D GPE (black solid line), 1D RGPE (blue dot-dashed line), and 1D NPSE (red dashed line) for the parameters in the experiment of Ref.~\cite{Engels}. Energy in units of $\hbar\omega_r$ where $\omega_r=2\pi\times203$~Hz.
\label{fig:1d_energy}}
\end{figure}

Using the RGPE and NPSE, we performed a series of reduced dimensionality simulations for otherwise the same set of parameters.  A comparison of the amount of energy transferred to the condensate as a function of the obstacle speed is shown in Fig.~\ref{fig:1d_energy}.  While both approaches have the same broad features, the quantitative comparison between the NPSE and the 3D GPE is most favourable and suggests that the NPSE captures the essential physics of the 3D simulations.

We note the presence of a small shoulder in the energy transfer curve below $v=0.7$~\mms\ for the NPSE and below  $v=1.3$~\mms\ for the RGPE that is not present in the 3D GPE. This suggests the 1D system may be more easily excited than the 3D system at low obstacle speeds. The assumption that the 3D condensate density at each point along the path of the obstacle can be well represented by a 1D reduction is less accurate when the obstacle is travelling through the low density regions at the extreme ends of the elongated condensate, where the density gradient is higher. This can significantly impact the energy transfer at low speed obstacles, which are only able to create excitations in these extremities.

% -- FIGURE 7 -- %
% -- FIGURE 7 -- %

Having found a reasonable 1D representation of the experiment, the next step is to establish a link to the 1D homogeneous system where the breakdown of superfluidity occurs at well-defined thresholds.

% --- SUB-SECTION: Homogeneous Systems --- %
\subsection{Moving impurities in a 1D homogeneous BEC}
The response of a 1D homogeneous condensate to a moving impurity has been well studied (see for example, Refs.~\cite{Hakim, Leboeuf, Watanabe}). It has been shown that superfluidity breaks down above a well-defined threshold impurity speed $v_{-}$, and this coincides with the onset of the emission of dark solitons and dispersive waves \cite{Leszczyszyn}. Above the threshold speed, the rate at which excitations are created increases steadily with impurity speed, before peaking, then decreasing, and eventually vanishing at a higher threshold speed $v_{+}$. 

Watanabe \emph{et al.} have performed a stability analysis of the 3D homogeneous GPE with an impurity that varied only in 1D~\cite{Watanabe}. The authors worked in the hydrodynamic regime where the impurity width is much larger than the local healing length, and assumed the instability occurs at the point when the local superfluid velocity was equal to the local sound velocity. Applying the breakdown condition to the Bernoulli equation leads to an expression for the critical current. For the present work, we have recast their result as
% EQUATION
\begin{equation}
\left( \frac{v}{\sqrt{2}c} \right)^2 - \left(\frac{3}{2}\right)^{\frac{2}{3}} \left( \frac{v}{\sqrt{2}c} \right)^{\frac{2}{3}} + 1 - \frac{U_0}{\mu} = 0,
\label{eq:watanabe_velocity}
\end{equation}
% EQUATION
where $U_0$ is the maximum amplitude of the impurity potential and $\mu = m c^2$ is the chemical potential. Physically, the smallest solution $v_-$ of Eq.~(\ref{eq:watanabe_velocity}) corresponds to the critical velocity, while the largest solution $v_+$ is the speed above which no energy is transferred to the system. These thresholds take into account the reduction in local density due to the potential height of the obstacle, as well as the increase in local fluid flow about the obstacle required by continuity. Comparison with  time-dependent 1D GPE simulations verifies that the stability thresholds predicted using Eq.~(\ref{eq:watanabe_velocity}) provide good estimates of the onset and cut-off velocities for the creation of solitons and other excitations of the system described earlier. Note that while our simulations are strongly in the hydrodynamic regime where the estimate is valid, the estimate only becomes exact in the hydrodynamic limit.

\begin{figure}
\includegraphics[width=0.82\columnwidth]{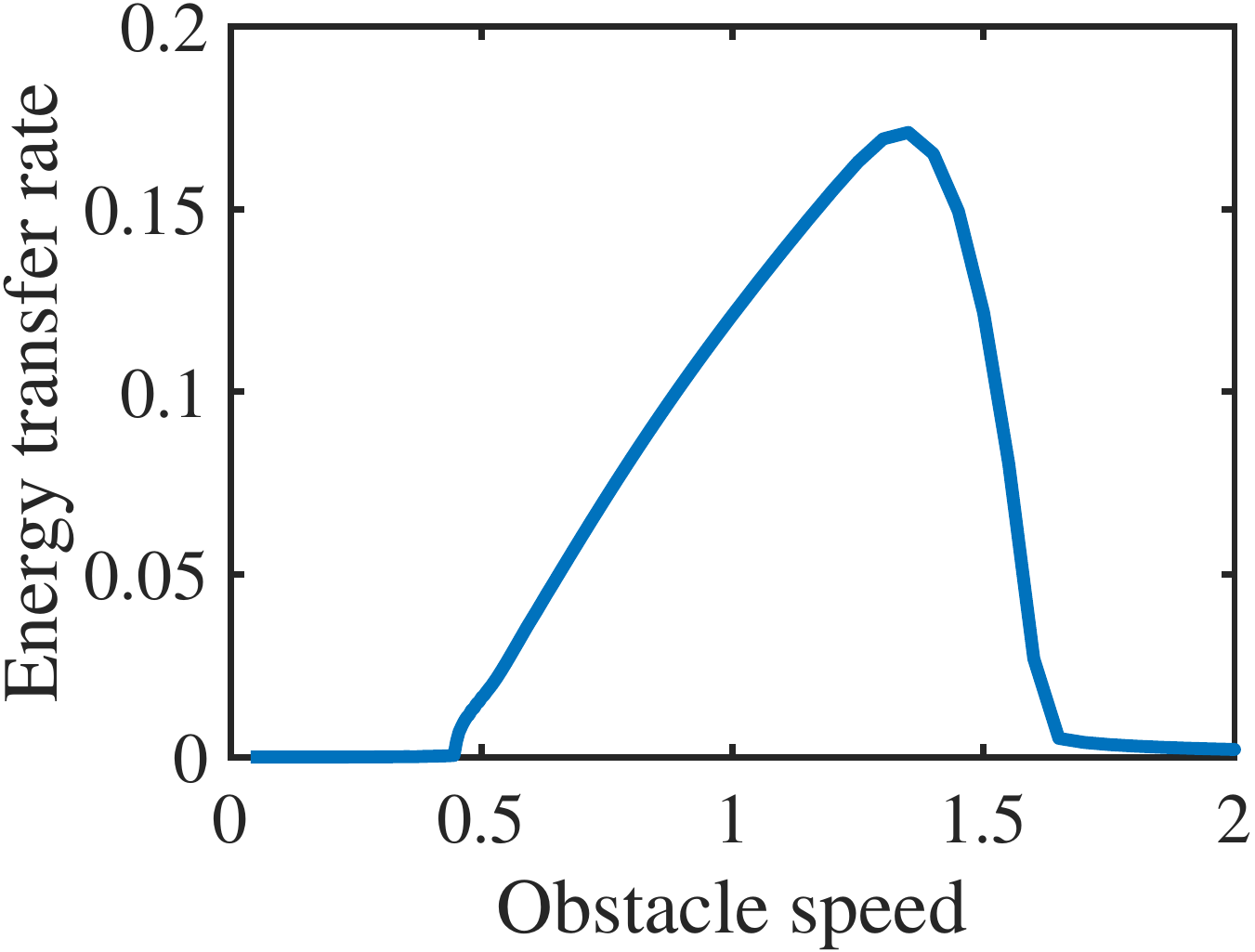}
\caption{Energy transfer for a 1D homogeneous system with $g=1,\mu=1$, a Gaussian obstacle with $U_0/\mu = 0.20$ and width $\sigma = 10$.  The upper and lower thresholds for energy transfer are  $v_-\approx0.45$ and $v_+\approx1.65$, consistent with predictions from Eq.~(\ref{eq:watanabe_velocity}) of $v_- = 0.47$ and $v_+ = 1.56$. The speed of sound in this system is $c=1$. All quantities in dimensionless units.
\label{fig:homogeneous_example}}
\end{figure}

 Past studies have generally focused on the critical thresholds and qualitative behaviour in between the thresholds. For our purposes, it is useful to also have some understanding of the behaviour between the thresholds, which can be accessed by looking at the energy transfer. We provide an example simulation of the energy transfer using illustrative parameters in Fig.~\ref{fig:homogeneous_example}.

% -- FIGURE 8 -- %
% -- FIGURE 8 -- %

% --- SUB-SECTION: Trapped Systems --- %
\subsection{Local density approximation for elongated harmonically trapped BECs}
Here we make use of the results from the previous section combined with the local density approximation to estimate the energy transfer for the experiment of Engels and Atherton~\cite{Engels}. The previous results can be applied in this situation as the experiment approximately falls in the hydrodynamic regime (i.e. the obstacle size ($\sim10$  $\mu$m) is larger than the local healing length ($\sim0.01-1$ $\mu$m) in the majority of the BEC).  We use a 1D approximation for the line density of the condensate ground state, and assume the rate of energy transfer to be a local function of the density, denoted as $R(x)$ where 
% EQUATION
\begin{equation}
R(x) = R(n(x)),
\end{equation}
% EQUATION
where $n(x)$ is the local density. This approach only captures the energy transfer that arises from the interaction of the impurity and the quiescent condensate, and ignores energy transfer due to interactions between the impurity and other excitations (for example, as may result from a collision between  the impurity and an excitation that was previously formed). 

The energy transfer rate $R(x)$ needs to be determined numerically. We do this by using the density $n(x)$ at each point $x$ along the path of the obstacle as input to a 1D homogeneous GPE simulation described by  
\begin{equation}
i \frac{\partial}{\partial t}\psi_h(y,t) = \left( -\frac{1}{2} \frac{\partial^2}{\partial y^2} + U(y,t) + g_{h}\left|\psi_h(y,t)\right|^2\right)\psi_h(y,t),
\label{eq:homogeneous_GP}
\end{equation}
where we have used $y$ as the spatial coordinate to avoid confusion with $x$, which refers to a particular position on the path of the obstacle. For clarity, parameters used in the homogeneous simulations are distinguished with a subscript $h$ (e.g.~$\mu_h$, $g_h$, etc). 

To replicate the local conditions, we choose $\psi_h(y,t=0) = \sqrt{n(x)}$. The potential $U_h(y)$ has the same form as Eq.~(\ref{eq:gaussian_obstacle}), with $U_0 = 0.24\mu$. Other parameters can also be determined from the density. 

For the RGPE, this is straightforward, $\mu_h = gn(x)$, $g_h=\mu_h/n(x)$ (equal to $g$) and $c_h=\sqrt{\mu_h}$ (equal to $\sqrt{gn}$), with the speed of the obstacle set to $v_h = [v/c(x)] c_h$ (equal to $v$), where $c(x)$ is the local speed of sound at the point $x$ on the path and $c_h$ is the speed of sound in the homogeneous system. 

When using the density profile obtained from the NPSE
\begin{equation}
\mu_h = \left\{ \frac{[3\alpha{n(x)}+2]^2}{4[\alpha{n(x)}+1]}\right\}^{1/2},
\end{equation}
 from which we can calculate $g_h = \mu_h/n(x)$, and 
 \begin{equation}
 c_h = \frac{1}{2}\left\{ \alpha{n(x)} \frac{4+3\alpha{n(x)}}{  [1+\alpha{n(x)}]^{3/2} }\right\}^{1/2}
 \end{equation}
  is used to calculate $v_h = [v/c(x)] c_h$.

% -- FIGURE 9 -- %
\begin{figure*}
{ \includegraphics[width=0.80\columnwidth]{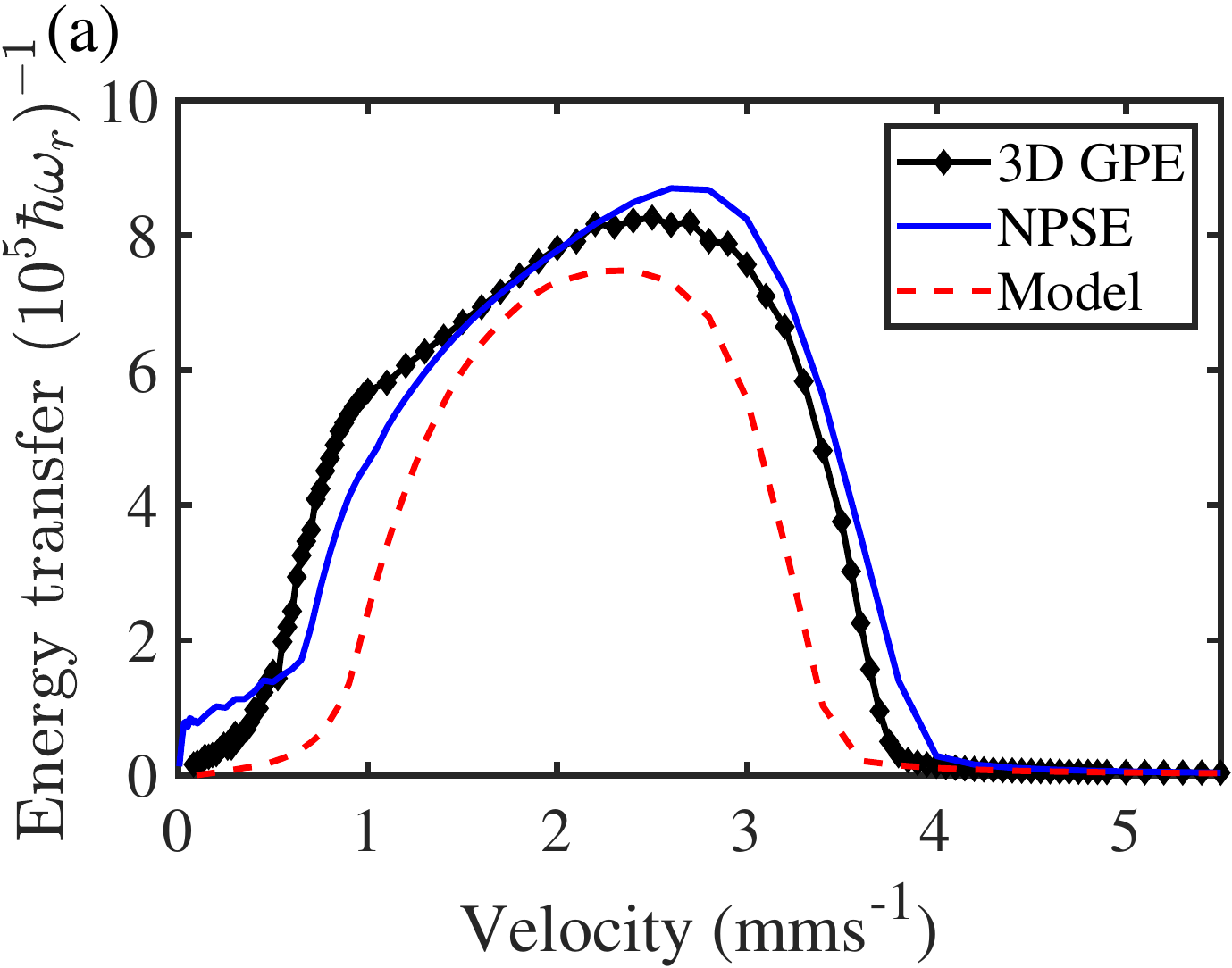} }\quad 
{ \includegraphics[width=0.80\columnwidth]{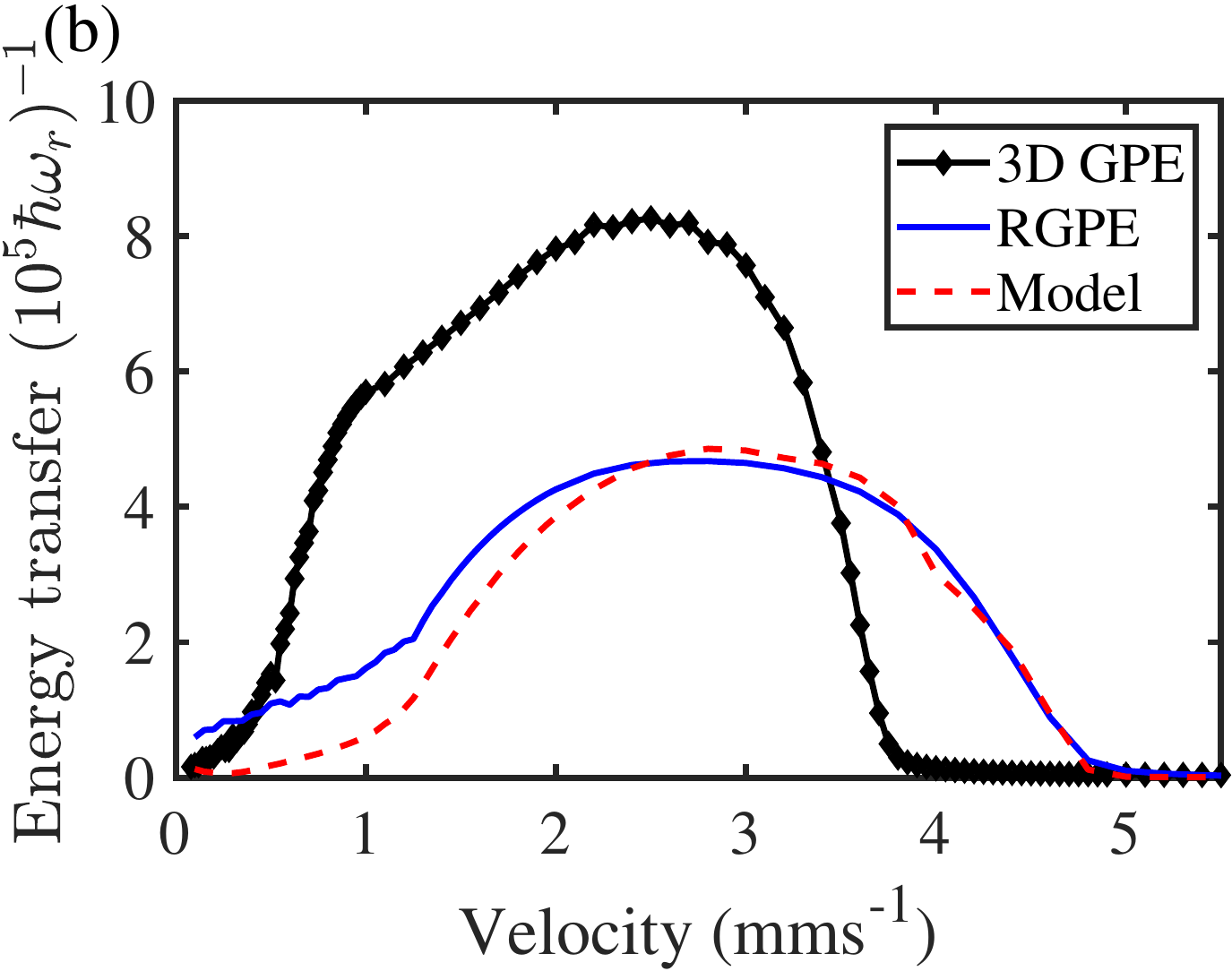} } \\
\qquad \\
{ \includegraphics[width=0.80\columnwidth]{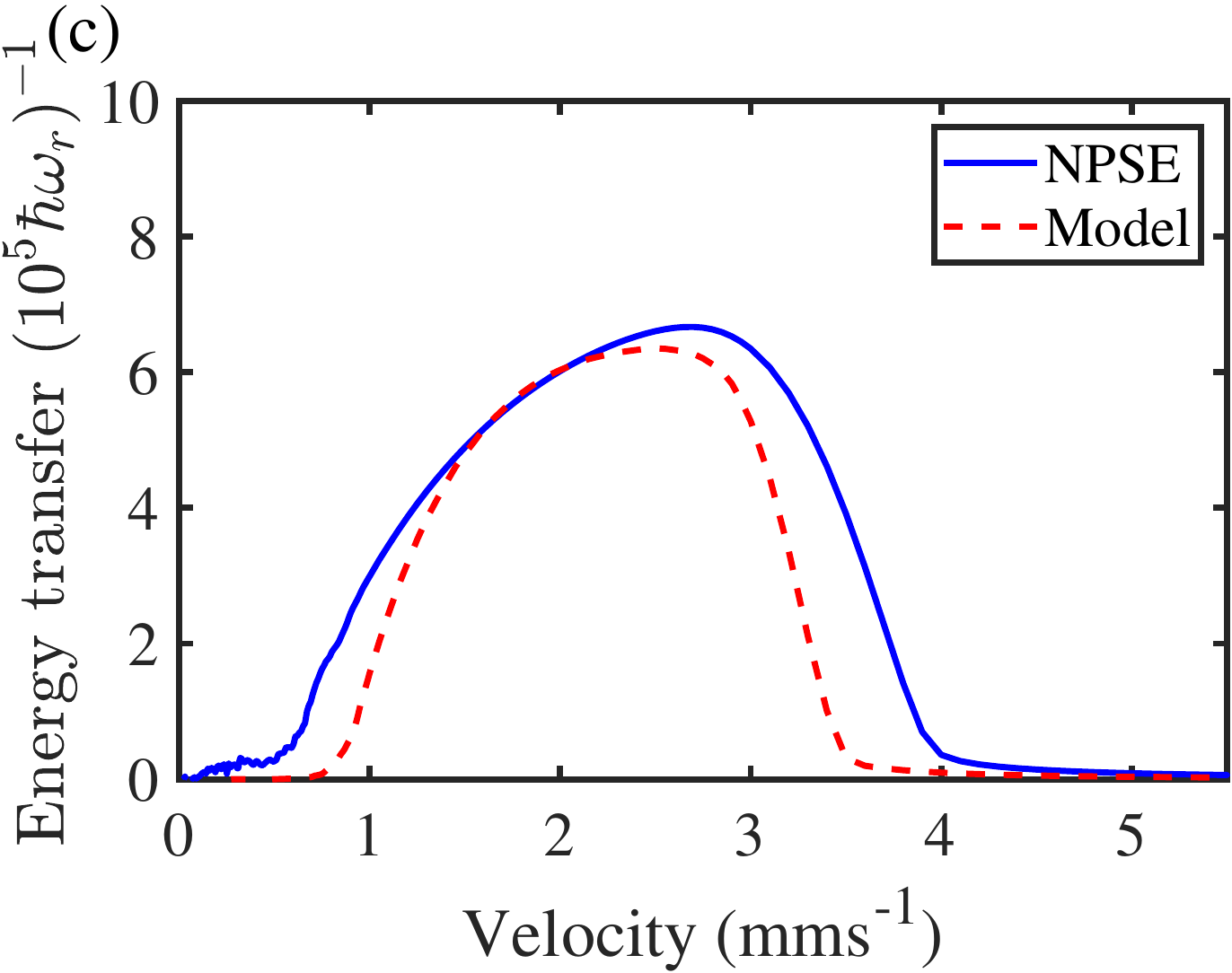} }\quad
{ \includegraphics[width=0.80\columnwidth]{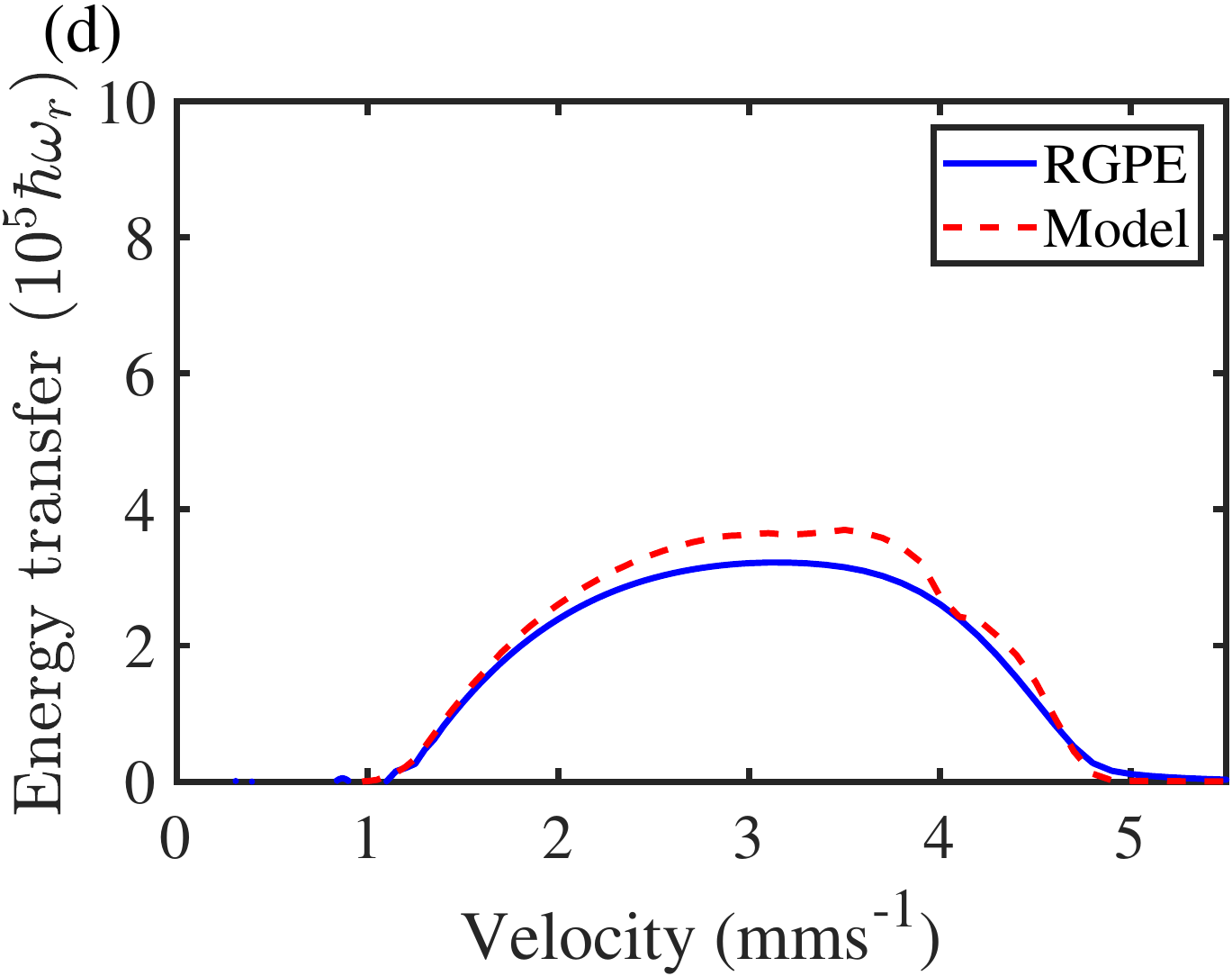} }
\caption{Total energy transfer as a function of obstacle velocity from numerical simulations (blue solid line) as compared to the LDA model  (red dashed line) as described in the text. Simulations using the 3D GPE are shown for reference (black line with solid dots). (a) The experimental obstacle path described in Ref.~\cite{Engels} for the NPSE.  (b) The experimental obstacle path described in Ref.~\cite{Engels} for the RGPE. (c) The truncated obstacle path as described in the text for the NPSE. (d) The truncated obstacle path as described in the text for the RGPE.
\label{fig:compare_LDM}}
\end{figure*}
% -- FIGURE 8 -- %

To calculate the rate of energy transfer in the simulations, the obstacle is suddenly turned on at time $t=0$ moving at constant speed $v_h$, then instantly turned off at time $t_f$. The average rate of energy transfer is estimated as
% EQUATION
\begin{equation}
R = \frac{E(t_f)-E(0)}{ t_f}.
\label{eq:rate_equation}
\end{equation}
% EQUATION
The energy transfer function $E(t)$ has step-like features due to the discrete soliton formation events, and so the calculation of the average rate $R$ has some dependence on the integration time $t_f$.  This sensitivity is particularly pronounced for velocities $v$ very close to $v_c$, when the soliton formation rate is low. However, it converges reasonably rapidly as $t_f$ is increased. We have ensured that our integration time is sufficiently long for our determination of $R$ to be accurate.

Having constructed the function $R(x)$, we can now calculate the energy transfer for the experiments of Engels and Atherton~\cite{Engels}. Noting $dt = dx/v$, the total energy transfer can be determined by
% EQUATION
\begin{eqnarray}
\Delta E(v) &=&\frac{1}{v} \int_{x_0}^{x_f}{ R[n(x),\mu(x);v/c(x))] \, dx},
\label{eq:energy_transfer_equation}
\end{eqnarray}
% EQUATION
where $x_0$ and $x_f$ are the initial and final positions of the obstacle trajectory.

% -- FIGURE 10 -- %  
\begin{figure*}
\includegraphics[width=0.94\textwidth]{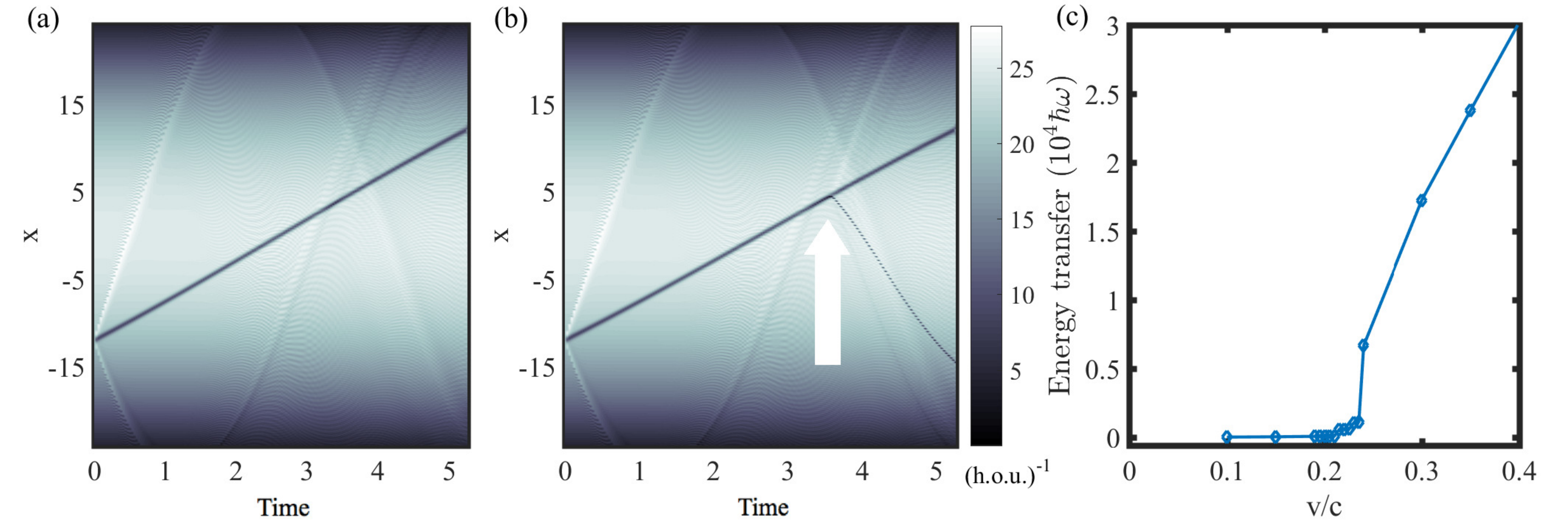}
\caption{Simulation of a critical velocity experiment in a harmonically trapped 1D BEC with an obstacle with fixed fractional speed, depth, and width. (a) and (b) show the 1D BEC density as a function of space and time and for an obstacle moving at $v/c = 0.210$ and  $v/c = 0.211$, respectively. The arrow in (b) indicates where a grey soliton is formed. (c) The energy transferred to the BEC in this scenario as a function of $v/c$.  The step change in behaviour and sharp transition in energy transfer at $v/c = 0.211$ is indicative of a clear critical velocity fraction. Parameters for the simulations are $g=14$ and $\mu=353$ in harmonic oscillator units.
\label{fig:1D_GPE_Simulations}}
\end{figure*}
% -- FIGURE 9 -- %

The energy transfer predicted by this model as a function of the obstacle velocity is compared to the direct integration of the NPSE and the RGPE in Fig.~\ref{fig:compare_LDM}(a) and Fig.~\ref{fig:compare_LDM}(b) respectively. The model largely reproduces the total energy transfer for both the NPSE and the RGPE, demonstrating that the experimental measurements can be interpreted as being due to the obstacle creating excitations  when its speeds exceeds a local density-dependent critical velocity equal to the local speed of sound.  We note that the model slightly underestimates the NPSE at lower and higher velocities, and the RGPE at lower velocities.  We believe this is at least in part because soliton formation, and hence energy transfer, occurs as discrete events in the simulations, but are approximated as occurring continuously in the model (as estimated by Eq.~(\ref{eq:rate_equation})). This difference is minimised when averaged over the entire path of an obstacle, but is more evident when the obstacle velocity is very small or very large, and only creates excitations in a narrow region.

In the Engels and Atherton experiment~\cite{Engels}, the density varies gradually (of order $\sim20-30\%$) in the central regions of the condensate, which accounts for two thirds of the path. At moderate velocities (i.e. between 1--3\mms) the obstacle produces excitations in the majority of this central region and the energy model is more accurate. At very low or high velocities, excitations only occur in low density ``edges'' of the condensate or in the highest density centre and the model is less reliable.

To test this conjecture we have performed further NPSE and RGPE simulations using the same parameters but with a modified obstacle path that begins approximately one quarter of the condensate length into the system and ends in the same location as in the experiment~\cite{Engels}, thus it avoids traversing the low density edges. By our conjecture, this should alleviate the discrepancy at lower velocities. The NPSE and RGPE energy transfers for this scenario are shown in Fig.~\ref{fig:compare_LDM}(c) and Fig.~\ref{fig:compare_LDM}(d) respectively, along with the model predictions. In this case there is better agreement between the model and the simulations at lower velocities, while agreement at higher velocities is largely unaffected. 

Overall, this straightforward model provides a reasonably accurate prediction of the energy transfer. We emphasise that while $R(x)$ is found numerically using the homogeneous GPE, its underlying excitation thresholds and mechanisms are well understood. This outcome demonstrates the excitation thresholds applying to the experiment can be quantitatively understood as the aggregate of local thresholds determined by the Landau criterion. Therefore, our modelling suggests that the results of Engels and Atherton~\cite{Engels} experiment are consistent with a critical velocity equal to the speed of sound, once suitably modified to take account of the variations in local density and flow conditions across the path of the moving obstacle.

% -- FIGURE 11 -- %
\begin{figure*}
\includegraphics[width=0.82\textwidth]{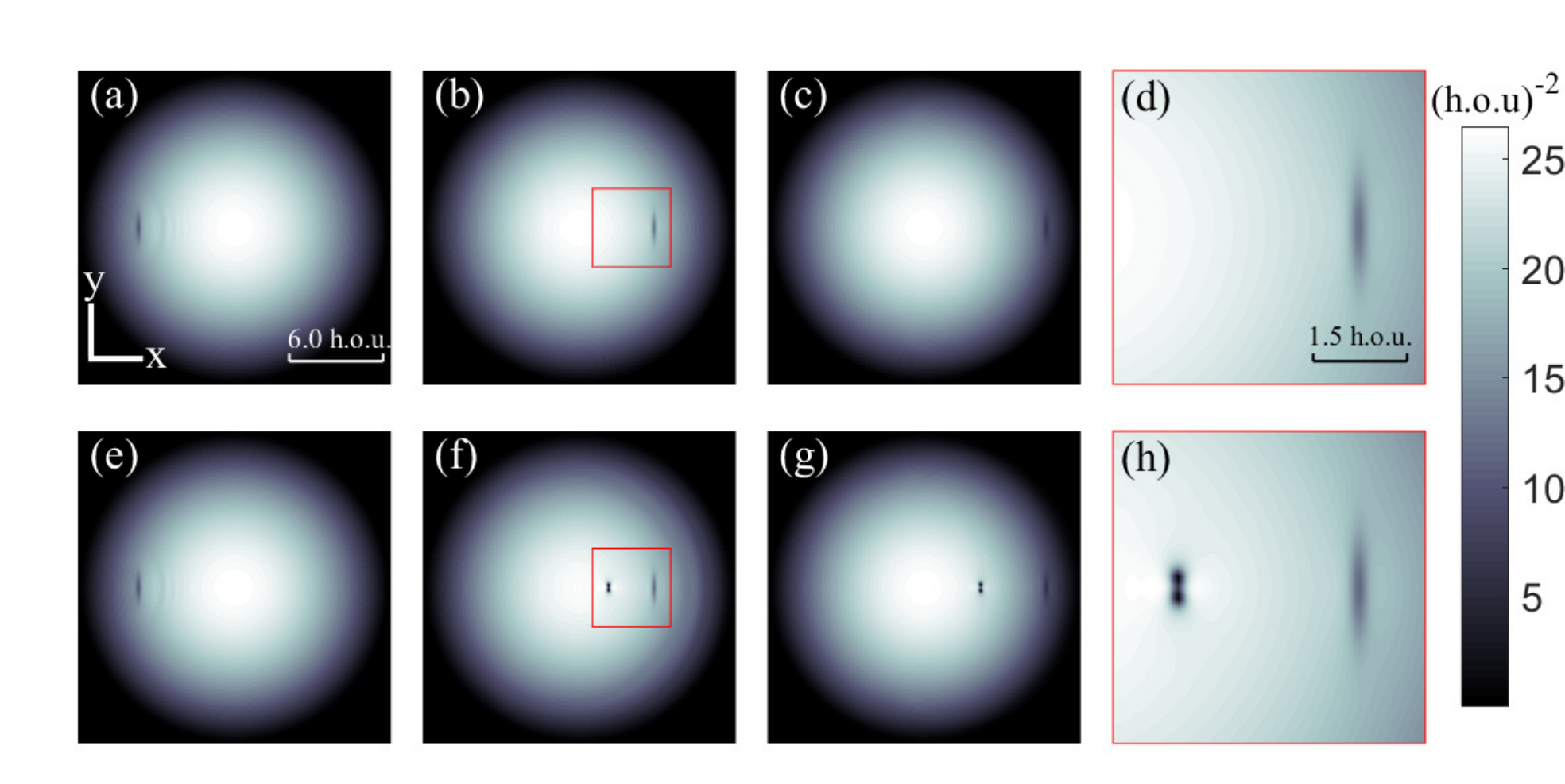}
\caption{Simulation of a critical velocity experiment in a harmonically trapped 2D BEC with an obstacle with fixed fractional speed, depth, and width. (a--c) BEC density at times $\omega t= 0.27, 1.9,$ and $2.4$  respectively for $v/c = 0.71$, which is below the fractional critical velocity.  (d) BEC density from the red box in (c) at higher magnification. (e--g) BEC density at  times $\omega t= 0.26, 1.8$, and $2.4$ respectively for $v/c = 0.72$, which is above the critical velocity fraction.  (h) BEC density from the red box in (f) at higher magnification, showing the presence of a vortex-antivortex pair. Simulation parameters are $g=2.67$ and $\mu=69.36$. The length scales are in harmonic oscillator units of $\sqrt{\hbar/m\omega}$.
 \label{fig:2D_GPE_Simulations}}
\end{figure*}
% -- FIGURE 11 -- %

% -- SECTION 5: FRACTIONAL APPROACH -- %
\section{Measuring the critical velocity in a harmonically trapped condensate}

Our results indicate that, within the validity of the local density approximation, the local speed of sound is indeed equal to the critical velocity for superfluidity in a trapped BEC according to the Gross-Pitaevskii equation.  Given the enduring interest in reaching a conclusive agreement between the predictions of the GPE regarding superfluidity and experiments on trapped BEC, it would be of interest to perform further experiments to quantitatively determine the critical velocity of weakly-interacting BECs.

One approach is to perform superfluidity experiments in homogeneous or near-homogeneous systems, which are becoming increasingly common in the laboratory~\cite{Gaunt,Chomaz,Gauthier}.  However, our results suggest another approach to critical velocity experiments with harmonically trapped BECs. Instead of an obstacle moving with a fixed speed, one could imagine an experiment where the obstacle width, depth, and speed are dynamically adjusted to remain at constant fractions of the local healing length, chemical potential, and speed of sound, respectively.  A second adjustment would be to avoid the low density outer regions of the BEC where the LDA is not valid~\cite{Carretero-Gonzalez}. In this scenario, for a given choice of obstacle parameters, one would expect to observe clear threshold behaviour in the energy transfer. Further, if the impurity parameters are chosen such that the system is in the hydrodynamic regime, one could expect to observe a critical velocity fraction $F_c$ consistent with solutions in $v/c$ to Eq.~(\ref{eq:watanabe_velocity}). 

Typical experimental parameters suggest that such an experiment could be readily performed. As an example, if using laser beams to generate the impurity, the width can be readily varied in the range 5 to 20 $\mu$m in a harmonically trapped condensate with local healing lengths of order $\xi=10$ to $\xi=100$~nm for $10^5$ atoms. The required variation in laser beam intensity to maintain the impurity depth at a set fraction of the chemical potential is approximately one order of magnitude, while the laser sweep speed required is within an achievable range of $v=0.1$~\mms to $v=1.0$~\mms.

% -- FIGURE 12 -- %
\begin{figure}
\includegraphics[width=0.9\columnwidth]{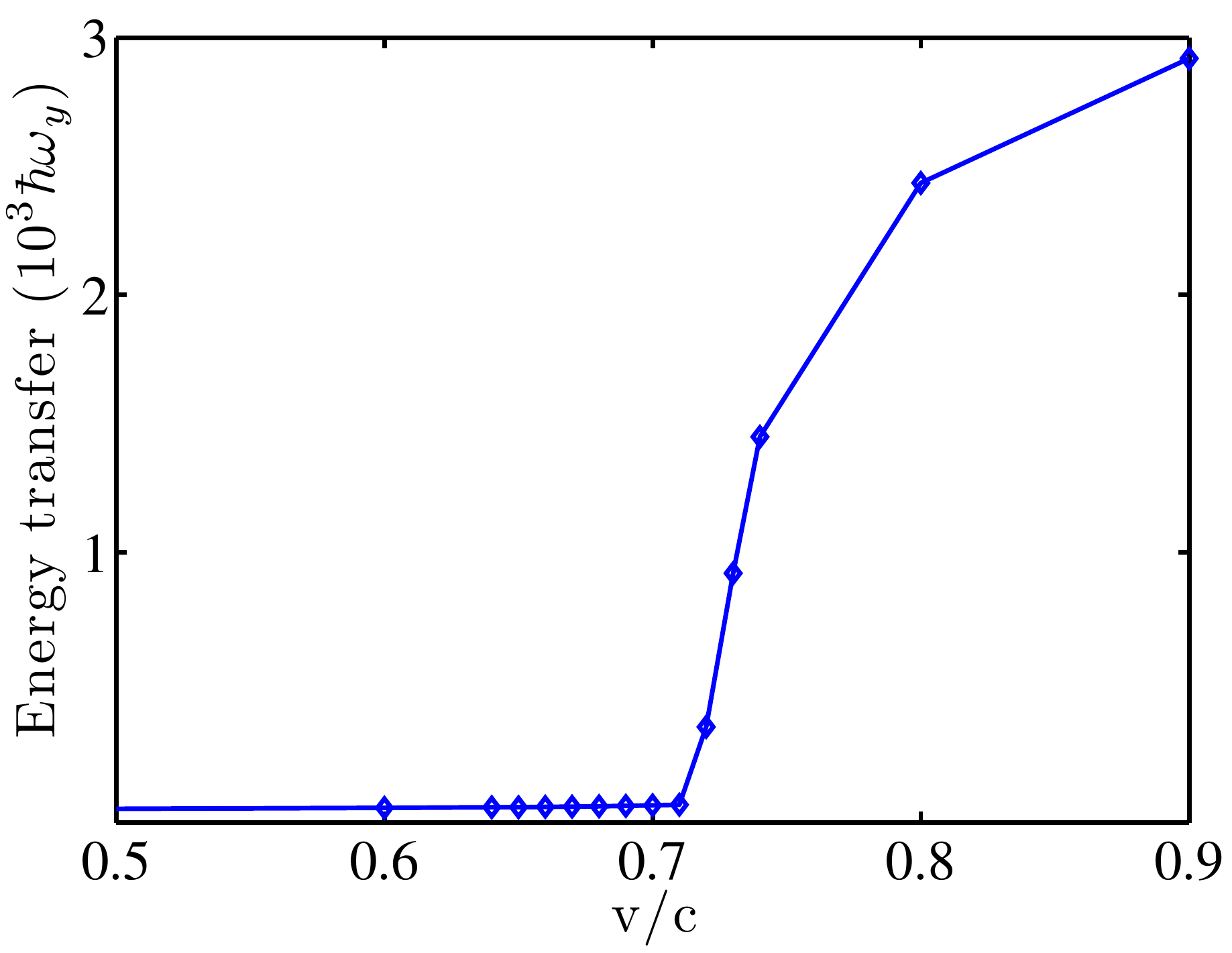}
\caption{Energy transfer for 2D GPE simulation with parameters $g=2.67$ and $\mu=69.36$. A sharp increase in energy transfer occurs above $v/c = 0.72$. 
 \label{fig:2D_GPE_Simulations_Energy}}
\end{figure}
% -- FIGURE 12 -- %

To illustrate this approach, we consider an archetypal experiment where a harmonically trapped BEC is stirred by a moving obstacle in similar fashion to Ref.~\cite{Engels}. For the purpose of demonstration, we use the simple 1D GPE described by Eq.~\ref{eq:RGP}. The obstacle is introduced at $x=-12$, and moved to $x=+12$ such that $v/c(x)=$ constant, with a relative obstacle depth continuously adjusted to $0.24\mu[n(x)]$ where $\mu[n(x)]$ is the local chemical potential. The obstacle width is fixed at eight times the local healing length. Figure~\ref{fig:1D_GPE_Simulations}(a) shows a simulation just below the critical velocity fraction at $v/c(x) = 0.210$, where there are no disturbances in the superfluid aside from some minor dispersive waves from the initial abrupt introduction of the obstacle. When the simulation is repeated with the obstacle speed increased to $v/c(x) = 0.211$,  a single grey soliton is emitted near the centre of the BEC as in Fig.~\ref{fig:1D_GPE_Simulations}(b). Figure~\ref{fig:1D_GPE_Simulations}(c) shows the energy transfer to the BEC as a function of the fraction of the speed of sound. A sharp increase can be seen at $v/c(x) = 0.211$, the critical fraction.

We find this behaviour extends to the two-dimensional case. We performed analogous simulations for the 2D GPE where the breakdown of superfluidity occurs by the formation of vortex-antivortex pairs~\cite{Neely, Jackson_Vortex}.  Example results are shown in Fig.~\ref{fig:2D_GPE_Simulations}, where the obstacle was a Gaussian potential with the same parameters as for the 1D simulation and with a fixed width of two harmonic oscillator units in the direction perpendicular to the motion.  Again we find that there is a fractional velocity for which there is a distinct breakdown of superfluidity, as can be seen in Fig.~\ref{fig:2D_GPE_Simulations_Energy}.

Interestingly, the critical velocity fraction predicted by Eq.~(\ref{eq:watanabe_velocity}) appears to approximately hold for 2D trapped condensates. Previously an indication that vortex pair creation is consistent with the Landau criterion has been reported in the work of Ref.~\cite{Crescimanno}, where estimates for the energy transfer and momentum change accompanying the creation of a single vortex pair was used to predict the critical velocity. 

\begin{figure}
\includegraphics[width=0.9\columnwidth]{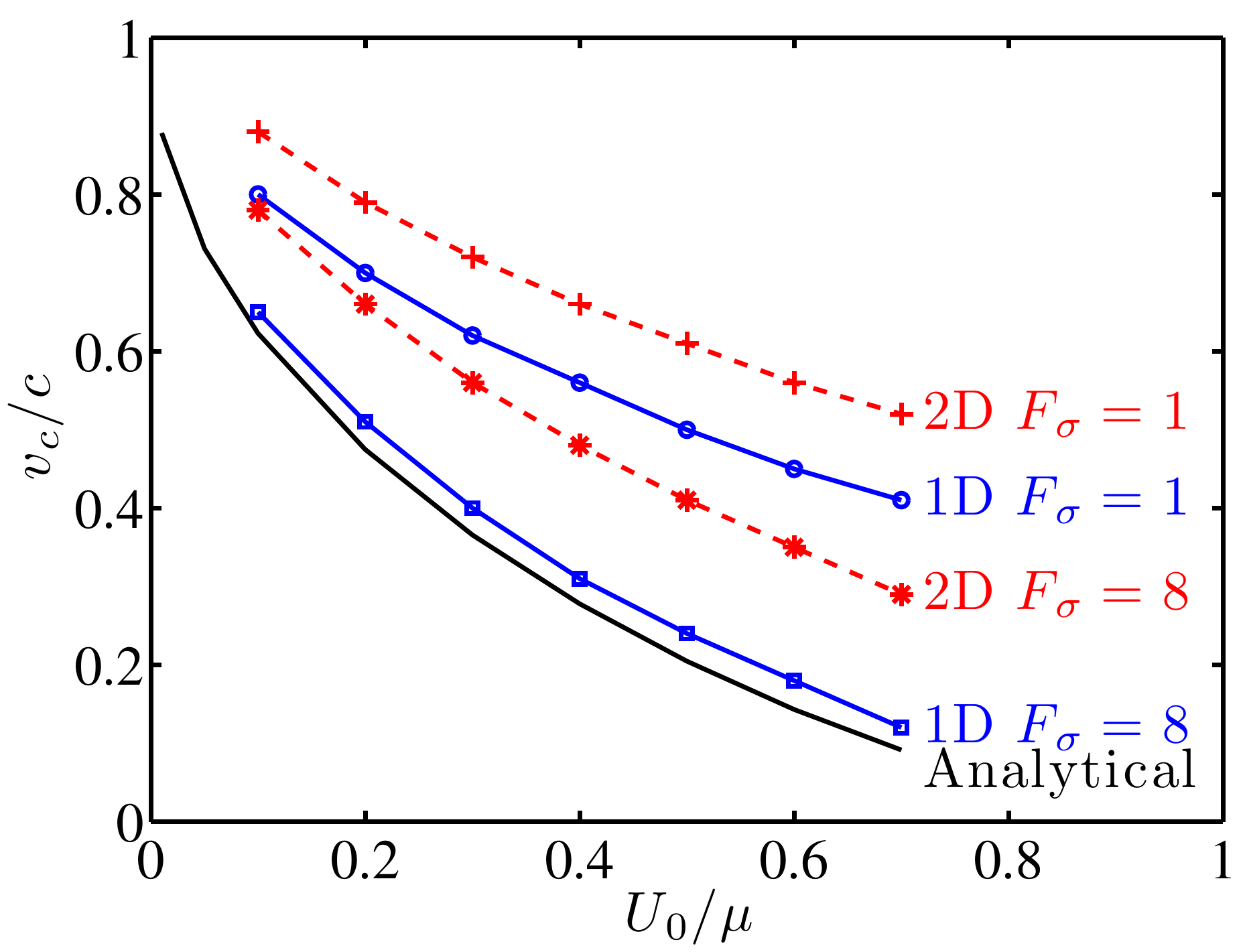}
\caption{Critical velocity fraction $v/c$ as a function of the obstacle depth $U_0/\mu$ for a narrow ($\sigma/\xi = 1$) and wide  ($\sigma/\xi = 8$) obstacle in 1D GPE simulations (blue circles and squares, respectively) and 2D GPE  simulations (red plus and asterisk symbols, respectively). Analytical predictions of Eq.~(\ref{eq:watanabe_velocity}) are shown as the solid black line~\cite{Watanabe}.
 \label{fig:fractional_thresholds}}
\end{figure}

Finally, we compare the fractional critical velocities  determined numerically from simulations of trapped BECs in 1D and 2D to analytical predictions by  Watanabe \emph{et al.}~\cite{Watanabe}. The analytical predictions are valid in the hydrodynamic limit $\sigma/\xi\gg1$~\cite{Pavloff} --- we have performed calculations for $\sigma/\xi = 1$ and $8$. In Fig.~\ref{fig:fractional_thresholds} we show the predicted and simulated fractional velocity as a function of the fractional obstacle depth $U_0/\mu$. As might be expected, we find that a greater obstacle depth reduces the critical velocity. For $\sigma/\xi \ll 1$ we expect the critical velocity to tend to $v_c/c \rightarrow 1$.

% -- FIGURE 13 -- %
% -- FIGURE 13 -- %

% -- SECTION 6: CONCLUSIONS -- %
\section{Conclusions}
In conclusion, we have performed a computational study of the experiments of Engels and Atherton~\cite{Engels} who analysed the excitations of an elongated, harmonically trapped BEC subjected to a localised potential moving through the condensate at constant speed.  By simulating the three-dimensional Gross-Pitaevskii equation we have obtained results that are consistent with their measurements, but which do not support an interpretation of a single critical velocity for this experiment~\cite{Engels}.

To better understand these results, we have constructed an effective 1D model of energy transfer to the system that allows us to measure the aggregate effect of an obstacle traversing regions of differing critical velocities. We have shown that, within the local density approximation, the experimental observations are consistent with with the existence of a local critical velocity equal to the local speed of sound, as modified by the local density and flow conditions.  In light of this model, we propose an experiment to measure the critical velocity for harmonically trapped BECs based on moving obstacles at a constant fraction of the local speed of sound within the bulk of the condensate.   This experimental  procedure should show a sharp increase in excitations of the BEC at a well-defined fraction of the local speed of sound, in agreement with theoretical predictions based on the Landau criterion.  Experiments based on this strategy would provide a simple manner to further probe superfluidity in trapped BECs, and will assist with the reconciliation of the previous discrepancies between the experimental and theoretical understanding of the Landau critical velocity in inhomogeneous condensates.\\

% --- SECTION: ACKNOWLEDGEMENTS -- %
\begin{acknowledgments}
The authors acknowledge many useful discussions with Tod Wright, and thank him for a critical reading of an early draft of the manuscript.  CF acknowledges discussions with John Close, and computational support from Graham Dennis. This research was supported by Australian Research Council Discovery Projects Nos. DP1094025 and DP110101047, and the Australian Research Council Centre of Excellence in Future Low-Energy Electronics Technologies (project number CE170100039) and funded by the Australian Government.
\end{acknowledgments}

%\bibliography{superfluid_bib}
%

\bibliographystyle{apsrev4-1}
\end{document}